\documentclass[prl, twocolumn,showpacs,preprintnumbers,amsmath,amssymb]{revtex4}
\usepackage{mathrsfs}

\usepackage{graphicx}
\usepackage{dcolumn}
\usepackage{amsmath}
\usepackage{epsfig}
\usepackage{upgreek}
\usepackage[]{subfigure}
\usepackage{color}
\usepackage[bookmarks=false]{hyperref}
 
\def\opex{ Opt.\ Express }

\begin{document}

\title{Unconditional Security of Time-energy Entanglement Quantum Key Distribution\\ using Dual-basis Interferometry}
\author{Zheshen Zhang}%
\email{zszhang@mit.edu}%
\author{Jacob Mower}
\author{Dirk Englund}
\author{Franco N. C. Wong}
\author{Jeffrey H. Shapiro}
\affiliation{Research Laboratory of Electronics, Massachusetts Institute of Technology\\
77 Massachusetts Avenue, Cambridge, Massachusetts 02139, USA}%
\date{\today}

\begin{abstract}
High-dimensional quantum key distribution (HDQKD) offers the possibility of high secure-key rate with high photon-information efficiency. We consider  HDQKD based on the time-energy entanglement produced by spontaneous parametric downconversion, and show that it is secure against collective attacks. Its security rests upon visibility data---obtained from Franson and conjugate-Franson interferometers---that probe photon-pair frequency correlations and arrival-time correlations.  From these measurements an upper bound can be established on the eavesdropper's Holevo information by translating the Gaussian-state security analysis for continuous-variable quantum key distribution so that it applies to our protocol. We show that visibility data from just the Franson interferometer provides a weaker, but nonetheless useful, secure-key rate lower bound. To handle multiple-pair emissions, we incorporate the decoy-state approach into our protocol. Our results show that over 200\,km transmission distance in optical fiber, time-energy entanglement HDQKD could permit a 700\,bit/sec secure-key rate, and a photon information efficiency of 2\,secure-key bits per photon coincidence in the key-generation phase using receivers with 15\% system efficiency.
\end{abstract}

\pacs{03.67.Dd, 03.67.-a, 42.50.Dv, 03.67.Hk}

\maketitle

Quantum key distribution (QKD)  \cite{gisin02} promises unconditionally secure communication by enabling one-time pad transmission between remote parties, Alice and Bob.  Continuous-variable QKD (CVQKD) \cite{grosshans03,lodewyck07} and discrete-variable QKD (DVQKD) \cite{bennett84,ekert91} utilize infinite-dimensional and finite-dimensional Hilbert spaces, respectively.  CVQKD exploits the wave nature of light to encode multiple bits into each transmission, but it has been limited to 80\,km in optical fiber \cite{lodewyck07,xuan09,jouguet13} because the eavesdropper (Eve) can obtain partial information from a beam-splitting attack.  The predominant DVQKD protocol is Bennett-Brassard 1984 (BB84), which uses a two-dimensional Hilbert space.  Decoy-state BB84 \cite{wang05,lo05} has demonstrated nonzero secure-key rates over 144\,km in free space \cite{schmitt-manderbach07} and 107\,km in optical fiber \cite{rosenberg07}, but its photon information efficiency cannot exceed 1\,key-bit/sifted-photon.  

High-dimensional QKD (HDQKD) using single photons \cite{zhang08} can utilize the best features of the continuous and discrete worlds, with the Hilbert space of single-photon arrival times providing an appealing candidate for its implementation.  The time-energy entanglement  of photon pairs produced by spontaneous parametric downconversion (SPDC) has been employed in HDQKD experiments \cite{tittel00, ali-khan07}, although these works lacked rigorous security proofs.  Security proofs for time-energy entangled HDQKD have been attempted by discretizing the continuous Hilbert space to permit use of DVQKD security analyses \cite{zhang08,nunn13}, but  the validity of the discretization approach has not been proven. CVQKD security analysis \cite{navascues06,garcia06} uses the quadrature-component covariance matrix to derive a lower bound on the secure-key rate in the presence of a collective attack.  We shall take an analogous approach---using the time-frequency covariance matrix (TFCM)---for our time-energy entanglement HDQKD protocol. 

The TFCM for our protocol can be obtained using the dispersive-optics scheme from \cite{mower12}, although dense wavelength-division multiplexing may be required to do so \cite{mower11}.  An experimentally simpler technique---utilizing a Franson interferometer---has been conjectured \cite{tittel00,ali-khan07} to be sufficient for security verification. Its robustness against some specific attacks has been discussed \cite{ali-khan07,brougham13}, but security against collective attacks has not been proven, and \cite{brougham13} suggests such a proof may be impossible. 

This Letter proves that time-energy entanglement HDQKD can be made secure against Eve's collective attack when a Franson interferometer is used for security verification in conjunction with a dispersion-based frequency-difference measurement.  Our proof relates the Franson interferometer's fringe visibility to the TFCM's frequency elements that, together with the frequency-difference measurement, establishes an upper bound on Eve's Holevo information. We introduce another nonlocal interferometer---the conjugate-Franson interferometer---and link its fringe visibility to the TFCM's arrival-time elements \cite{grice10}.  Employing both interferometers increases the secure-key rate. 

Our fringe visibility results presume that the entanglement source emits at most one photon-pair in a measurement frame, which need not be the case for SPDC.  Thus we incorporate decoy-state operation \cite{wang05,lo05} to handle multiple-pair emissions. We will show that time-energy entanglement HDQKD could permit a 700\,bit/sec secure-key rate over 200\,km transmission distance in optical fiber.  We will also show that a photon information efficiency of 2\,secure-key bits per photon coincidence can be achieved in the key-generation phase using receivers with 15\% system efficiency.  Before beginning our security analysis, we provide a brief explanation of our protocol.  

Suppose Alice has a repetitively-pumped, frequency-degenerate SPDC source that, within a time frame of duration $T_f$\,sec which is centered at time $t_m = 3mT_f$, emits a single photon-pair in the state \cite{footnote0}
\begin{eqnarray}
\lefteqn{\hspace*{-.1in}|\psi_m\rangle_{SI} \propto} \nonumber \\[.05in] 
&&\hspace*{-.25in}\int\!dt_S\!\int\!dt_I\, e^{-(t_+ - t_m)^2/4\sigma^2_{\rm coh}-t_-^2/4\sigma^2_{\rm cor}-i\omega_Pt_+}|t_S\rangle_S|t_I\rangle_I
\label{GaussBiphoton}
\end{eqnarray}
for some integer $m$. In this expression: $\omega_P$ is the pump frequency; $|t_S\rangle_S$ ($|t_I\rangle_I$) represents a single photon of the signal (idler) at time $t_S$ ($t_I$); $t_+ \equiv (t_S+t_I)/2$; $t_- \equiv t_S-t_I$; the root-mean-square coherence time $\sigma_{\rm coh}= T_f/\sqrt{8\ln(2)} \sim$\,ns is set by the pump pulse's duration; and the root-mean square correlation time $\sigma_{\rm cor} = \sqrt{2\ln(2)}/2\pi B_{\rm PM}\sim$\,ps is set by the reciprocal of the full-width at half-maximum (FWHM) phase-matching bandwidth, $B_{\rm PM}$, in Hz.  Now suppose that, despite propagation losses and detector inefficiencies, Alice and Bob detect the signal and idler, respectively, from the preceding photon pair and record the associated arrival times \cite{footnote1,footnote2}.  After many such frames, they use public communication to reconcile their arrival-time data, resulting in their sharing $n_R$ random bits per post-selected frame, i.e., frames used for key generation in which Alice and Bob both made detections.  How many of those bits are secure against Eve's collective attack?  Before turning to the security analysis, we pause for a brief note about Eq.~(\ref{GaussBiphoton}).  This expression is an oft-used approximation for the post-selected biphoton state produced by an SPDC source, see, e.g., \cite{ali-khan07}.  Moreover, entanglement engineering can be employed to achieve a close match to a truly Gaussian biphoton wave function \cite{Dixon13}.   

Our security analysis begins with the positive-frequency field operators, $\hat{E}_S(t)$ and $\hat{E}_I(t)$, for the linearly-polarized single spatial-mode signal and idler fields emitted by Alice's source, and their associated frequency decompositions:
\begin{subequations}
\begin{align}
\hat{E}_S(t) &= \int \frac{d\omega}{2\pi}\hat{A}_S(\omega)e^{-i(\omega_P/2+\omega)t}\\
\hat{E}_I(t) &=\int \frac{d\omega}{2\pi}\hat{A}_I(\omega)e^{-i(\omega_P/2-\omega)t}.
\end{align}
\end{subequations}
The time-domain field operators $\hat{E}_S(t)$ and $\hat{E}_I(t)$ annihilate signal and idler photons, respectively, at time $t$, and they obey the canonical commutation relations, $[\hat{E}_J(t),\hat{E}_K^\dagger(u)] = \delta_{JK}\delta(t-u)$, for $J,K = S,I$.  Their frequency-domain counterparts, $\hat{A}_S(\omega)$ and $\hat{A}_I(\omega)$, annihilate signal and idler photons at detunings $\omega$ and $-\omega$, respectively.  Our interest, however, is in the arrival-time and angular-frequency operators, 
\begin{subequations}
\label{eqDefOperator}
\begin{align}
\hat{t}_J &= \int\!dt\, t\hat{E}^\dag_J(t)\hat{E}_J(t),\\
\hat{\omega}_J &=  \int\!\frac{d\omega}{2\pi}\, \omega\hat{A}^\dag_J(\omega)\hat{A}_J(\omega),
\end{align}
\end{subequations}
for $J=S,I$, when only one photon-pair is emitted by the source.  Restricting these time and frequency operators to the single-pair Hilbert space implies that they measure the arrival times and frequency detunings of the signal and idler photons.  It also leads to the commutation relation $[\hat{\omega}_J,\hat{t}_K] = i\epsilon_J\delta_{JK}$ \cite{suppl}, where $\epsilon_S = - \epsilon_I = 1$, making these operators conjugate observables analogous to the quadrature components employed in CVQKD, and justifying our translating CVQKD's covariance-based security analysis \cite{navascues06,garcia06} to our protocol.  

To exploit the connection to CVQKD, we define an observable vector $\hat{\boldsymbol{\mathcal{O}}} = [\begin{array}{cccc}\hat{t}_S & \hat{\omega}_S & \hat{t}_I & \hat{\omega}_I\end{array}]^T$. For a single-pair state, the mean value of $\hat{\boldsymbol{\mathcal{O}}}$ is $\textbf{m} = \langle\hat{\boldsymbol{\mathcal{O}}}\rangle$, and  the TFCM is $\boldsymbol{\Gamma} = \langle(\Delta\hat{\boldsymbol{\mathcal{O}}}\Delta\hat{\boldsymbol{\mathcal{O}}}^\dagger + {\rm h.c.})\rangle/2$, where $\Delta\hat{\boldsymbol{\mathcal{O}}} \equiv \hat{\boldsymbol{\mathcal{O}}}-\textbf{m}$ and ${\rm h.c.}$ denotes Hermitian conjugate. The characteristic function associated with the single-pair state is $ \chi({\boldsymbol\zeta}) = \langle e^{i{\boldsymbol\zeta}^T \hat{\boldsymbol{\mathcal{O}}}}\rangle$. Given the covariance matrix $\boldsymbol{\Gamma}$, the Gaussian state with $\chi({\boldsymbol\zeta}) = e^{i\boldsymbol{\zeta}^T\textbf{m} - {\boldsymbol\zeta}^T{\boldsymbol\Gamma}{\boldsymbol\zeta}/2}$ yields an ${\bf m}$-independent upper bound on Eve's Holevo information \cite{navascues06,garcia06,wolf06} when the SPDC source emits a single-pair state.

\begin{figure}[htb]
\centering
\subfigure{\includegraphics[width=3.25in]{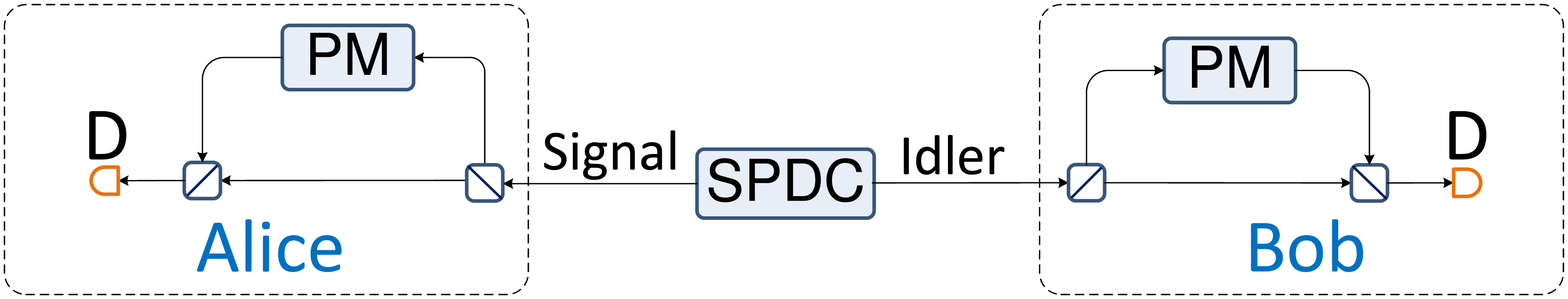}}
\subfigure{\includegraphics[width=3.25in]{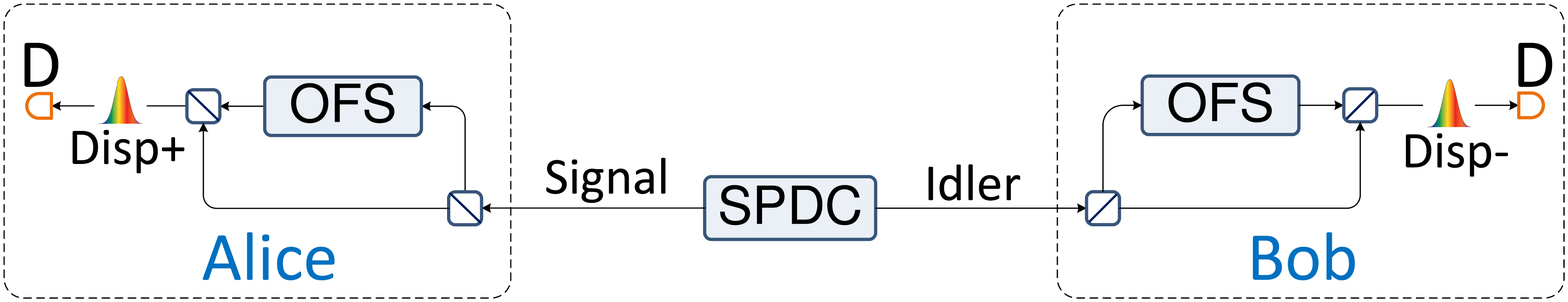}}
\caption{\label{figFranson} (color online) Top panel:  diagram for the Franson interferometer. Bottom panel: diagram for the conjugate-Franson interferometer. PM:  phase modulator; OFS: optical-frequency shifter; Disp$+$: positive dispersion element; Disp$-$: negative dispersion element. D: detector.}
\end{figure}

A direct, complete measurement of the TFCM is quite challenging, so we will resort to indirect measurements---using a Franson interferometer and a conjugate-Franson interferometer---that provide useful partial information.  A Franson interferometer \cite{franson89}, shown in the top panel of Fig.~\ref{figFranson}, consists of two unequal path-length Mach-Zehnder interferometers, with the signal going through one and the idler going through the other. The time delay $\Delta T$ between each Mach-Zehnder's long and short paths is much greater than the correlation time $\sigma_{\rm cor}$, ruling out local interference in the individual interferometers. It is also greater than the FWHM detector timing jitter, $\delta T$, so that coincidences are only registered when both photons go through the long or the short path. Each long path is equipped with a phase modulator, imparting phase shifts $ e^{-i\phi_S} $ and $ e^{-i\phi_I} $ to the signal and the idler respectively.  The following Lemma shows that Franson measurements, augmented by dispersion-based frequency measurements, bound the signal-idler frequency correlations \cite{suppl}:\\
\textit{Lemma 1:} For a single-pair state let $V_{\rm FI}(\Delta T) = [P_{C_{\rm FI}}(0)-P_{C_{\rm FI}}(\pi)]/[P_{C_{\rm FI}}(0)+P_{C_{\rm FI}}(\pi)]$, where $P_{C_{\rm FI}}(\phi_S+\phi_I)$ is Alice and Bob's coincidence probability, be the $0$-$\pi$ fringe visibility when the Franson interferometer has delay $\Delta T$.  Then the variance of the signal-idler frequency difference satisfies
\begin{equation}
\label{eqVF}
\left<(\Delta\hat{\omega}_S-\Delta\hat{\omega}_I)^2\right> \leq \frac{2[1-V_\text{FI}(\Delta T)]}{\Delta T^2}+\frac{\langle(\widetilde{\omega}_S-\widetilde{\omega}_I)^4\rangle}{12}\Delta T^2,
\end{equation}
where $ \widetilde{\omega}_S $ ($ \widetilde{\omega}_I $) is the random variable associated with the measured signal (idler) angular frequency from the conjugate-Franson interferometer with its frequency-shifted arms disabled, i.e., when dispersion enables frequency correlations to be measured from arrival-time coincidences. 

A conjugate-Franson interferometer, shown in the bottom panel of Fig.~\ref{figFranson}, consists of two equal path-length Mach-Zehnders with one arm of each containing an electro-optic optical-frequency shifter. To rule out local interference, these devices shift the signal and idler frequencies by $-\Delta\Omega$ and $\Delta\Omega$, respectively, while phase modulators (not shown) apply phase shifts $e^{-i\phi_S}$ and $e^{-i\phi_I}$, as was done in the Franson interferometer.  The positive and negative dispersion elements have coefficients $\pm \beta_2$ satisfying $ \beta_2\Delta \Omega = \sqrt{2}\,T_g > \delta T $, where $T_g$ is the duration of detectors' coincidence gate \cite{footnote3}.  They time-disperse the signal and idler's frequency components so that two detectors suffice to measure their frequency coincidences \cite{mower12,suppl}.  The following Lemma shows that conjugate-Franson measurements, augmented by arrival-time measurements, bound the signal-idler arrival-time correlations \cite{suppl}:\\
\textit{Lemma 2:} For a single-pair state let $V_{\rm CFI}(\Delta \Omega) = [P_{C_{\rm CFI}}(0)-P_{C_{\rm CFI}}(\pi)]/[P_{C_{\rm CFI}}(0)+P_{C_{\rm CFI}}(\pi)]$, where $P_{C_{\rm CFI}}(\phi_S+\phi_I)$ is Alice and Bob's coincidence probability, be the $0$-$\pi$ fringe visibility when the conjugate-Franson interferometer has frequency shift $\Delta \Omega$.  Then the variance of the signal-idler arrival-time difference satisfies
\begin{equation}
\label{eqVC}
\left<(\Delta\hat{t}_S-\Delta\hat{t}_I)^2\right> \leq \frac{2[1-V_\text{CFI}(\Delta\Omega)]}{\Delta \Omega^2}+\frac{\langle(\widetilde{t}_S-\widetilde{t}_I)^4\rangle}{12}\Delta \Omega^2,
\end{equation}
where $ \widetilde{t}_S $ ($ \widetilde{t}_I $) is the random variable associated with the measured signal (idler) arrival time from the Franson interferometer with its long arms disabled.

Lemmas~1 and 2 will be used below to bound Eve's Holevo information for a frame in which Alice's source emits a single photon-pair.  Because there is no security assurance for multiple-pair emissions, we follow the lead of DVQKD by employing decoy states \cite{wang05,lo05}  to deal with this problem. In particular, Alice operates her SPDC source at several different pump powers enabling Bob and her to estimate the fraction, $F$, of their coincidences that originated from single-pair emissions \cite{footnote4}. 

To put an upper bound on Eve's Holevo information, we start from the following points:  (1) Symmetry dictates that only 10 TFCM elements need to be found.  Of these, $ \langle \Delta\hat{\omega}_S ^2\rangle $ and $ \langle \Delta\hat{t}_S^2\rangle $ are immune to Eve's attack because Eve does not have access to Alice's apparatus, which contains the SPDC source.  (2) Given the Franson and conjugate-Franson's fringe visibilities, making $ \langle \Delta\hat{t}_J \Delta\hat{\omega}_K\rangle \neq 0$, for $J,K = S,I$, does not increase Eve's Holevo information \cite{suppl}.  (3) From Lemmas~1 and 2 we can determine upper bounds on the excess noise factors $1+\xi_\omega \equiv \langle (\Delta\hat{\omega}_S-\Delta\hat{\omega}_I)^2\rangle/\langle (\Delta\hat{\omega}_{S_0}-\Delta\hat{\omega}_{I_0})^2\rangle$ and $1+\xi_t\equiv  \langle(\Delta\hat{t}_S-\Delta\hat{t}_I)^2\rangle/\langle (\Delta\hat{t}_{S_0}-\Delta\hat{t}_{I_0})^2 \rangle$, where $  \langle (\Delta\hat{t}_{S_0}-\Delta\hat{t}_{I_0})^2\rangle $  and $ \langle (\Delta\hat{\omega}_{S_0}-\Delta\hat{\omega}_{I_0})^2\rangle $ are the source's variances as measured by Alice during her source-characterization phase.
  
Points (1)-(3) specify a set, $ \mathcal{M} $, of physically allowed TFCMs that preserve the Heisenberg uncertainty relations for the elements of $\hat{\boldsymbol{\mathcal{O}}}$ which are implied by $[\hat{\omega}_J,\hat{t}_K] = i\epsilon_J\delta_{JK}$.  For each TFCM ${\boldsymbol\Gamma} \in \mathcal{M}$, the Gaussian state $\chi({\boldsymbol\zeta}) = e^{- {\boldsymbol\zeta}^T{\boldsymbol\Gamma}{\boldsymbol\zeta}/2}$ affords Eve the maximum Holevo information  \cite{wolf06,navascues06,garcia06}.  Using $\chi_{\boldsymbol{\Gamma}}(A;E)$ to denote that Holevo information, our partial information about $\boldsymbol{\Gamma}$ gives us the upper bound $\chi_{\xi_t,\xi_\omega}^\text{UB}(A;E) = \sup_{{\boldsymbol\Gamma} \in \mathcal{M}} [\chi_{\boldsymbol{\Gamma}}(A;E)]$ on what Eve can learn from a collective attack on a single-pair frame.  Thus Alice and Bob's secure-key rate (in bits/sec) has the lower bound \cite{lo05,gottesman04,suppl}
\begin{equation}
\text{SKR} \geq \frac{q p_r}{3T_f}\bigg[\beta I(A;B)-(1-F)n_R - F\chi^\text{UB}_{\xi_t,\xi_\omega}(A;E)\bigg].
\end{equation}
Here: $q$ is the fraction of the frames used for key generation (as opposed to Franson or conjugate-Franson operation or decoy-state transmission for parameter estimation); $p_r$ is the probability of registering a coincidence in a frame; $\beta$ is the reconciliation efficiency; and $I(A;B)$ is Alice and Bob's Shannon information.  

\begin{figure}
\centering
\subfigure{\includegraphics[width=1.6in]{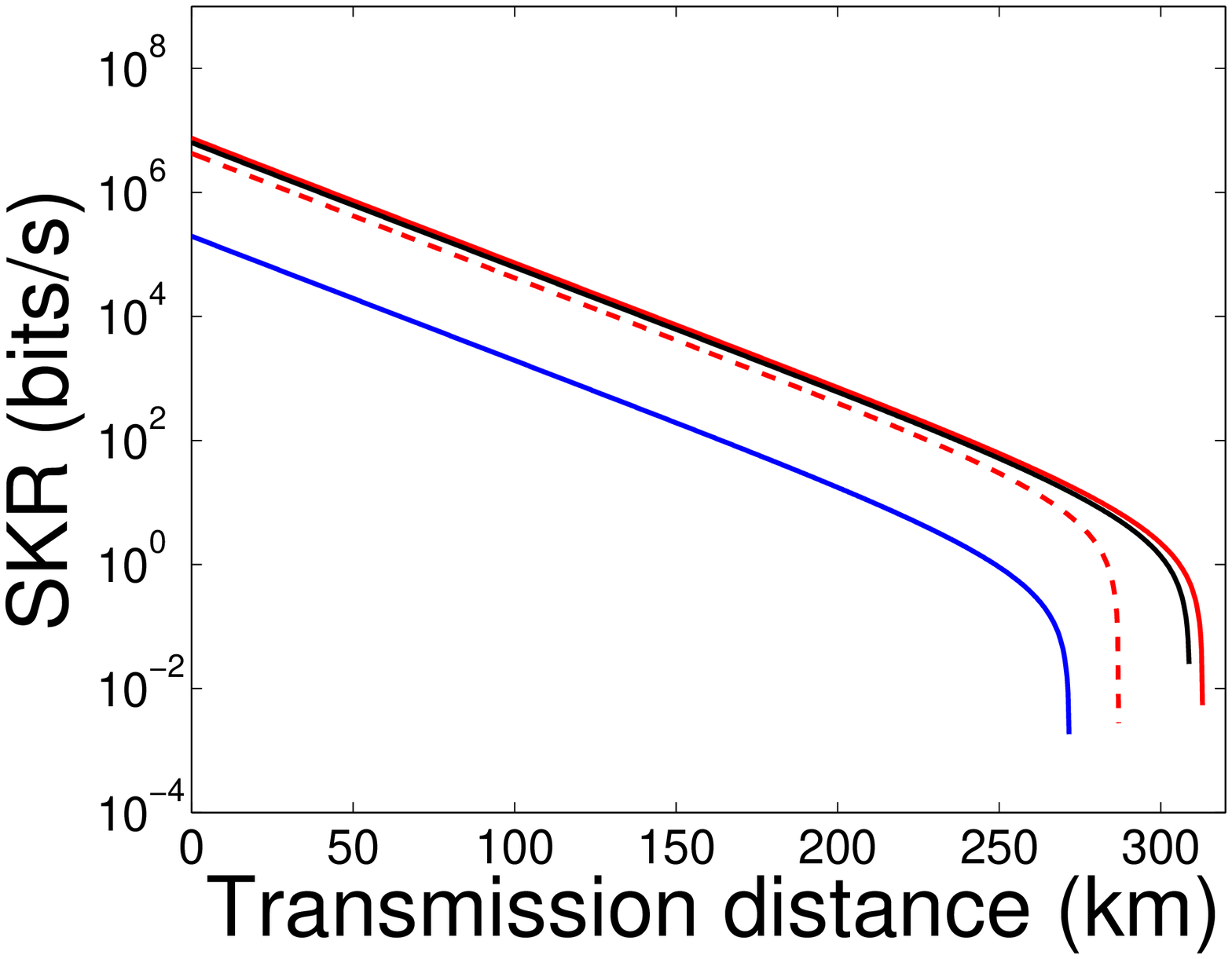}}
\subfigure{\includegraphics[width=1.6in]{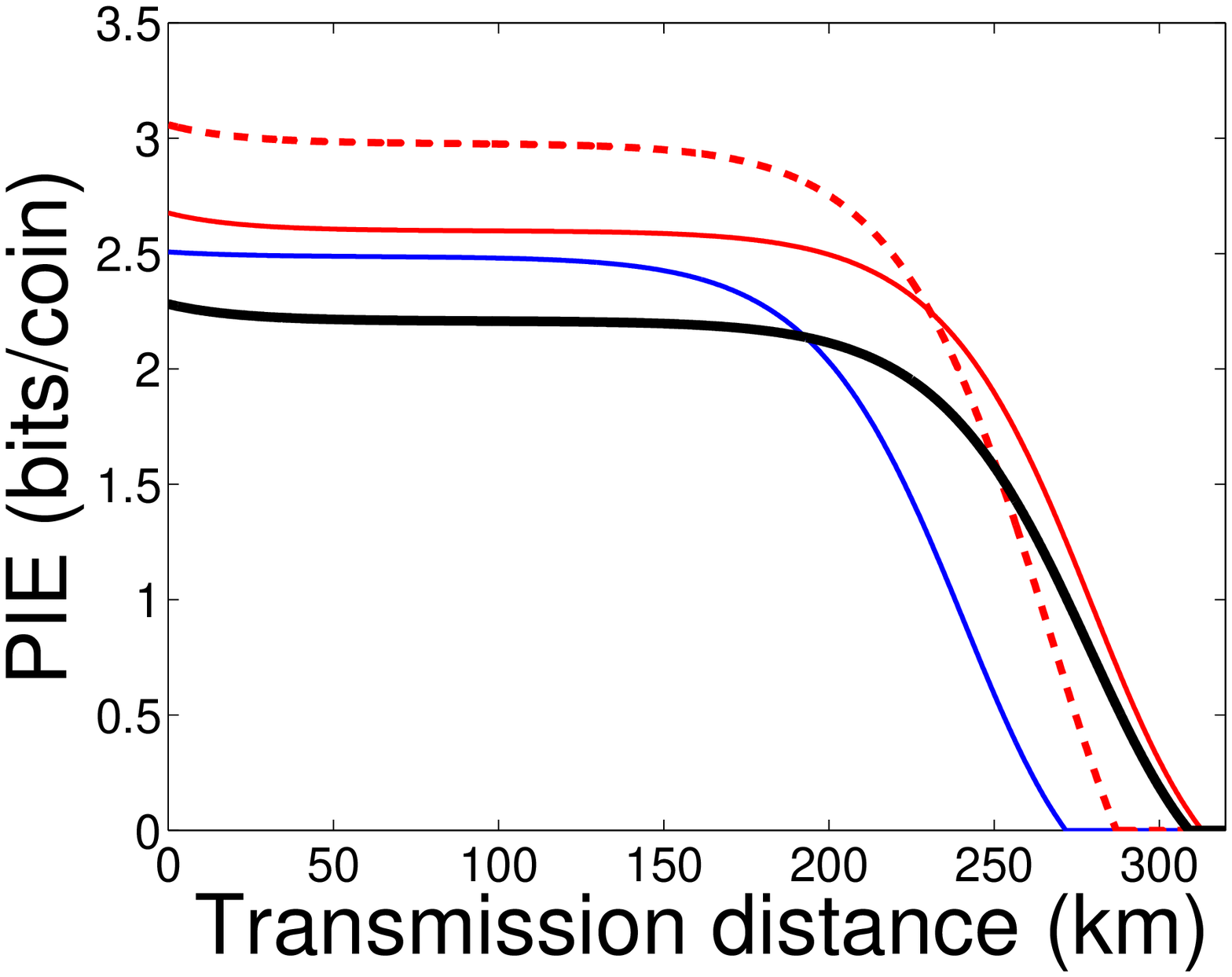}}
\caption{\label{figSKR_DeltaI_Distance} Left panel:  Alice and Bob's secure-key rate  versus transmission distance. Right panel: Alice and Bob's photon information efficiency versus transmission distance. Both assume the following values.  Entangled-pair flux: 0.01\,pairs/frame; detector timing jitter $ \delta T = 30 $\,ps; $B_{\rm PM} = 200\,$GHz; $\Delta T/\sqrt{2} = T_g = 3.6\,\delta T$; $\Delta\Omega/2\pi = 5\,$GHz; $\beta_2\Delta\Omega =  \sqrt{2}\,T_g$; $q = 0.5$; $\beta = 0.9$; $n_R = 8$; dark-count rate =  $10^3$/sec;  and fiber loss = 0.2\,dB/km. Solid curves:  $T_f = 16\,\delta T$ and $\xi_\omega = 0.22$. Dashed curves: $T_f = 32\,\delta T$ and $\xi_\omega = 1.01$. Blue curves: $\eta_A = \eta_B = 15\%$.  Red and black curves: $ \eta_A = \eta_B = 90\%$. Red and blue curves: $\xi_t = 41.5$. Black curves: $\xi_t = 400$.}
\end{figure}

Figure~\ref{figSKR_DeltaI_Distance}'s left panel plots Alice and Bob's secure-key rate (SKR) versus transmission distance for two frame durations and two system efficiencies for Alice ($\eta_A$) and Bob's ($\eta_B$) receivers, which use superconducting nanowire single-photon detectors (SNSPDs). To calculate the $ \xi_\omega $'s we assume that the measured $V_{\rm FI}$ values are their ideal values---93.25\% for $T_f = 16\,\delta T$ and 98.27\% for $T_f=32\,\delta T$---multiplied by 0.995.  (These $V_{\rm FI}$ values are achievable, see \cite{zhong13} in which 99.6\% fringe visibility was reported.)   For the red and blue curves, we calculate the $ \xi_t $'s by assuming that the measured $V_{\rm CFI}$'s are their ideal values---99.96\% for both $T_f = 16\,\delta T$ and $T_f = 32\,\delta T$---multiplied by 0.995.  For the black curve, $ \xi_t $ represents jitter-limited raw arrival-time measurements.
We see that QKD is possible out to 200\,km when Alice and Bob have receivers with 15\% system efficiency.  Going to 90\% system efficiency allows QKD out to 300\,km and increases the secure-key rate by nearly two orders of magnitude. 

There is an important point to make about the secure-key rate curves associated with the two $\xi_t$ values we have employed.  Constraining Eve to $\xi_t = 41.5$ requires the use of a conjugate-Franson interferometer, because jitter-limited raw arrival-time measurements cannot measure finer than $\xi_t = 400$ with our system parameters.  Surprisingly, $\xi_t = 400$ still yields a positive secure-key rate. This is because eavesdropping in one basis disturbs correlation in the conjugate basis. In our protocol, Alice and Bob generate key from the time basis, so degradation in the timing correlation does not increase Eve's Holevo information, although it slightly reduces Alice and Bob's mutual information and hence their secure-key rate.

The photon information efficiency (PIE) is defined to be the number of secure-key bits per photon coincidence in the key-generation phase, $\text{PIE} = \text{SKR}\,3T_f/qp_r$. The right panel of Fig.~\ref{figSKR_DeltaI_Distance} plots photon information efficiency versus transmission distance.  It shows that Alice and Bob achieve $\text{PIE} \ge 2$\,secure-bits/coincidence in the key-generation phase out to 200\,km when their receivers have 15\% system efficiency.

Our  protocol sacrifices potential secure-key rate when detector timing jitter, $\delta T$, exceeds the SPDC source's correlation time, $\sigma_{\rm cor}$, i.e., $I(A;B)$ cannot approach its ultimate limit of $\log_2(\sigma_{\rm coh}/\sigma_{\rm cor})$ bits/coincidence that is set by the source's Schmidt number.  That limit can be achieved with wavelength-division multiplexing (WDM) that makes the two-photon correlation time in each WDM channel comparable to the detector timing jitter \cite{mower11} and deriving key from time-frequency coincidences.  In this case, the conjugate-Franson interferometer becomes crucial, because part of the secure key information is obtained by frequency measurements. Nevertheless, the TFCM is still sufficient to bound Eve's Holevo information.

Before concluding, it behooves us to compare our security predictions with the individual-attack results reported in Brougham \em et al\/\rm. \cite{brougham13}.  The comparison is not entirely straightforward, because those authors considered a time-binned version of time-energy entanglement HDQKD with no multiple-pair emissions or dark counts, whereas our protocol operates in continuous time and includes both of those effects.  Consider the 1024-bin example from \cite{brougham13}, in which Eve obtains 6\,bits out of 10 when the Franson's fringe visibility is 99.2\% and 5\,bits when that visibility is 99.8\%.  To compare our results with those, we set $T_f = 1024\sqrt{2}\,\delta T$, so that Alice and Bob's mutual information equals 10\,bits/coincidence in the presence of $\delta T$ timing jitter when there are neither dark counts nor multiple-pair emissions.  Under these conditions, our security analysis sets upper bounds of 6.07\,bits and 5.83\,bits on Eve's Holevo information for 99.2\% and 99.8\% FI visibility.  

In summary, we adapted the Gaussian-state security analysis for CVQKD to our time-energy entanglement HDQKD protocol.  We showed that a Franson interferometer's fringe visibility suffices against arbitrary collective attacks when that measurement is used in conjunction with decoy states, which allow the fraction of single-pair SPDC frames to be estimated.  Adding a conjugate-Franson interferometer to the system enables tighter constraints on the TFCM, leading to a higher secure-key rate.  Our protocol promises QKD over 200\,km and multiple secure bits per coincidence. 

We thank T. Zhong for valuable discussions. This work was supported by the DARPA Information in a Photon Program through Army Research Office Grant No.\ W911NF-10-1-0416.

\widetext

\begin{center}{\Large\bf Supplemental Material}
\end{center}%

\section{Quasimonochromatic Signal and Idler Field Operators}
We have taken the positive-frequency field operators $\hat{E}_S(t)$ and $\hat{E}_I(t)$ for the signal and idler outputs of Alice's spontaneous parametric downconverter to be quasimonochromatic, photon-units operators \cite{Blow,YuenShapiro,Shapiro}.  As such this gives them the delta-function commutator, $[\hat{E}_J(t),\hat{E}_K^\dagger(u)] = \delta_{JK}\delta(t-u)$, for $J,K = S,I$.  Our security proof relies on time and frequency operators whose own delta-function commutator, proved in the next section, depends on the quasimonochromatic condition's validity.  The 200\,GHz phase-matching bandwidth assumed in the main text's system example corresponds to 0.1\% fractional bandwidth at the 1.55\,$\upmu$m fiber telecom wavelength.  Thus we are justified in assuming that our field operators obey the quasimonochromatic condition.

\section{Proof of $[\hat{\omega}_J,\hat{t}_K] = i\epsilon_J\delta_{JK}$ for $J,K=S,I$}
The operators for arrival-time and frequency-detuning measurements of the signal ($J=S$) and idler ($J=I$) are defined as follows
\begin{subequations}
\label{eq_Def_operators}
\begin{align}
\hat{t}_J &= \int\!dt\, t \hat{E}_J^\dag(t)\hat{E}_J(t)\\
\hat{\omega}_J &= \int\!\frac{d\omega}{2\pi}\, \omega \hat{A}_J^\dag(\omega)\hat{A}_J(\omega),
\end{align}
\end{subequations}
where the field operators are restricted to the Hilbert space spanned by the vacuum state and the single-photon time-domain states $\{\,|t\rangle_J : -\infty < t <\infty\,\}$, or, equivalently, the vacuum state and the single-photon frequency-domain states $\{\,|\omega\rangle_J : -\infty < \omega < \infty\,\}$ \cite{footnote1}.   In this Hilbert space the field operators reduce to 
\begin{equation}
\hat{E}_J(t) = |0\rangle_J{}_J\langle t|,
\label{Esingle}
\end{equation}
and 
\begin{equation}
\hat{A}_J(\omega) = |0\rangle_J{}_J\langle \omega|.
\label{Asingle}
\end{equation}

With the preceding restricted expressions for the field operators in time and frequency, we can easily derive the commutation relation $[\hat{\omega}_J,\hat{t}_K] = i\epsilon_J\delta_{JK}$ for $J,K=S,I$.  For $J\neq K$, the frequency-time commutator vanishes because $\hat{E}_J(t)$ commutes with $\hat{E}_K(t)$ and $\hat{E}_K^\dagger(t)$ and hence so too does $\hat{E}_J(t)$ with $\hat{A}_K(\omega)$ and $\hat{A}_K^\dagger(\omega)$.  For $J=K=S$ we have that
\begin{align}
\hat{\omega}_S\hat{t}_S - \hat{t}_S\hat{\omega}_S &= 
\int\!dt\,\int\!\frac{d\omega}{2\pi}\,\omega t[({}_S\langle \omega|t\rangle_S)|\omega\rangle_S{}_S\langle t| -({}_S\langle t|\omega\rangle_S)|t\rangle_S{}_S\langle \omega| ]
\\ 
&= \int\!dt\,\int\!\frac{d\omega}{2\pi}\,\omega t[e^{i\omega t}|\omega\rangle_S{}_S\langle t| - 
e^{-i\omega t}|t\rangle_S{}_S\langle \omega|],
\label{comm1}
\end{align}
where the second equality follows from 
\begin{equation}
|\omega\rangle_S = \int\!dt\,e^{-i\omega t}|t\rangle_S
\label{freqstateS}
\end{equation}
and
\begin{equation}
{}_S\langle t_1|t_2\rangle_S = \delta(t_1-t_2).
\label{comm2}
\end{equation}
Using Eqs.~(\ref{comm1}) and (\ref{comm2}), we then obtain the time-domain matrix elements
\begin{equation}
{}_S\langle t_1|[\hat{\omega}_S,\hat{t}_S]|t_2\rangle_S = (t_2-t_1)\int\!\frac{d\omega}{2\pi}\,\omega e^{-i\omega (t_1-t_2)}.
\label{matrixelement}
\end{equation}

Now, let $f(\tau)$ be any square-integrable function on $-\infty < \tau < \infty$ that is continuous at $\tau = 0$,  and has Fourier transform
\begin{equation}
F(\omega) = \int\!d\tau\,f(\tau)e^{-i\omega\tau}.
\end{equation}
We then have that
\begin{equation}
-\int\!d\tau\int\!\frac{d\omega}{2\pi}\,f(\tau)\tau \omega e^{-i\omega\tau} = 
-i\int\!\frac{d\omega}{2\pi}\,\omega\frac{dF(\omega)}{d\omega} = -i\frac{\omega}{2\pi} F(\omega)\vert_{-\infty}^\infty + i\int\!\frac{d\omega}{2\pi}\,F(\omega) =  i\int\!\frac{d\omega}{2\pi}\,F(\omega) = if(0),
\end{equation}
where the first equality is a Fourier-transform property, the second uses integration by parts, the third results from $f(\tau)$'s being square-integrable, and the fourth by inverse Fourier transformation.  Applying this result to Eq.~(\ref{matrixelement}) yields
\begin{equation}
{}_S\langle t_1|[\hat{\omega}_S,\hat{t}_S]|t_2\rangle_S  = i\delta(t_1-t_2),
\end{equation}
which is equivalent to $[\hat{\omega}_S,\hat{t}_S] = i$, thus completing our proof for $J=K=S$.  For $J=K=I$ the proof follows the same steps using
\begin{equation}
|\omega\rangle_I = \int\!dt\,e^{i\omega t}|t\rangle_I
\label{freqstateI}
\end{equation}
instead of Eq.~(\ref{freqstateS}).

\section{Proof of Lemma~1}
The coincidence probability for our Franson interferometer when only a single photon-pair has been emitted by Alice's source is \cite{footnote2}
\begin{align}
P_{C_{\rm FI}}(\phi_S,\phi_I) &= \frac{\eta}{16} \int\!dt\,\int_{t-T_g/2}^{t+T_g/2}\!du\, \left\langle\left[\hat{E}^\dag_S(t)+e^{i\phi_S}\hat{E}^\dag_S(t-\Delta T)\right]\left[\hat{E}^\dag_I(u)+e^{i\phi_I}\hat{E}^\dag_I(u-\Delta T)\right]\right.\nonumber\\[.05in]
&\times  \left.\left[\hat{E}_S(t)+e^{-i\phi_S}\hat{E}_S(t-\Delta T)\right]\left[\hat{E}_I(u)+e^{-i\phi_I}\hat{E}_I(u-\Delta T)\right]\right\rangle,
\end{align}
where $\eta$ accounts for propagation losses \cite{footnote3} and detection efficiencies, and $T_g > \delta T$, with $\delta T$ being the detectors' full-width at half-maximum (FWHM) timing jitter, is the duration of the coincidence gate.  Because only a single photon-pair has been emitted, we can employ Eq.~(\ref{Esingle}) and rewrite the coincidence probability as 
\begin{align}
P_{C_{\rm FI}}(\phi_S,\phi_I) &= \frac{\eta}{16} \int\!dt\,\int_{t-T_g/2}^{t+T_g/2}\!du\, \left\langle\left[|t\rangle_S{}_S\langle 0|+e^{i\phi_S}|t-\Delta T\rangle_S{}_S\langle 0|\right]\left[|u\rangle_I{}_I\langle 0|+e^{i\phi_I}|u-\Delta T\rangle_I{}_I\langle 0|\right]\right.\nonumber\\[.05in]
&\times  \left.\left[|0\rangle_S{}_S\langle t|+e^{-i\phi_S}|0\rangle_S{}_S\langle t-\Delta T|\right]\left[|0\rangle_I{}_I\langle u|+e^{-i\phi_I}|0\rangle_I{}_I\langle u-\Delta T|\right]\right\rangle.
\label{PcFIsingle}
\end{align}
Multiplying out the bracketed expressions inside the averaging leads to a sum of sixteen terms, but only four of them make nonzero contributions when $\Delta T > T_g > \delta T \gg \sigma_{\rm cor}$.  To see that this is so, let us first take the undisturbed (no eavesdropping) biphoton wave function \cite{footnote4}
\begin{equation}
\psi_{SI}(t_S,t_I) =  \frac{e^{-(t_S+t_I)^2/16\sigma^2_{\rm coh}-(t_S-t_I)^2/4\sigma^2_{\rm cor}-i\omega_P(t_S+t_I)/2}}{\sqrt{2\pi\sigma_{\rm coh}\sigma_{\rm cor}}}.
\label{biphotonTime}
\end{equation}
Averaging a term from Eq.~(\ref{PcFIsingle}) that contains $|t_A\rangle_S{}_S\langle t_B|\otimes|u_A\rangle_I{}_I\langle u_B|$ yields a result that contains $\psi_{SI}^*(t_A,u_A)\psi_{SI}(t_B,u_B)$.  Thus, unless $|t_A-u_A| \le T_g/2$ \em and\/\rm\ $|t_B-u_B|\le T_g/2$, this term will make a negligible contribution to $P_{C_{\rm FI}}$.  Indeed, the magnitude of each of these noncontributor terms is at least $\exp(-\Delta T^2/8\sigma^2_{\rm cor})$ times smaller than the terms we will retain.  For our paper's system example, which assumes $\Delta T = 152.7$\,ps and $\sigma_{\rm cor} = 0.937$\,ps, this attenuation factor is $\exp(-3320)$.  These numbers apply to detectors without timing jitter.  With timing jitter, the preceding attenuation factor becomes $\exp(-\Delta T^2\ln(2)/2\delta T^2)$, which, because $\delta T = 30\,$ps in our system example, equals $\exp(-8.98)$.  Now suppose there is eavesdropping.  Then, unless Eve's intrusion makes the root-mean-square arrival-time difference exceed the coincidence gate, such terms will still fail to contribute to the coincidence probability.  An Eve who makes that strong a disturbance will easily be detected and accounted for during reconciliation, so we will neglect that possibility in what follows, as we evaluate the coincidence probability.

After eliminating the twelve noncontributors to $P_{C_{\rm FI}}$, we get
\begin{align}
P_{C_{\rm FI}}(\phi_S,\phi_I) &=  \frac{\eta}{16} \int\!dt\,\int_{t-T_g/2}^{t+T_g/2}\!du\, \left\langle\left[|t\rangle_S|u\rangle_I{}_I\langle u|{}_S\langle t| + e^{-i(\phi_S + \phi_I)}|t\rangle_S|u\rangle_I{}_I\langle u-\Delta T|{}_S\langle t-\Delta T|\right.\right. \nonumber \\[.05in]
&+ \left.\left.e^{i(\phi_S+\phi_I)}|t-\Delta T\rangle_S|u-\Delta T\rangle_I{}_I\langle u|{}_S\langle t| + 
|t-\Delta T\rangle_S|u-\Delta T\rangle_I{}_I\langle u-\Delta T|{}_S\langle t-\Delta T|\right]\right\rangle.
\label{PcFIreduced}
\end{align}
At this point the integration limits for $u$ can be extended to $-\infty < u < \infty$, because the integrand vanishes for $|t-u| > T_g/2$ by an argument similar to what we just gave to go from Eq.~(\ref{PcFIsingle}) to Eq.~(\ref{PcFIreduced}).  Thus, we can write
\begin{align}
P_{C_{\rm FI}}(\phi_S,\phi_I) &=  \frac{\eta}{16} \int\!dt\,\int\!du\, \left\langle\left[|t\rangle_S|u\rangle_I{}_I\langle u|{}_S\langle t| + e^{-i(\phi_S + \phi_I)}|t\rangle_S|u\rangle_I{}_I\langle u-\Delta T|{}_S\langle t-\Delta T|\right.\right. \nonumber \\[.05in]
&+ \left.\left.e^{i(\phi_S+\phi_I)}|t-\Delta T\rangle_S|u-\Delta T\rangle_I{}_I\langle u|{}_S\langle t| + 
|t-\Delta T\rangle_S|u-\Delta T\rangle_I{}_I\langle u-\Delta T|{}_S\langle t-\Delta T|\right]\right\rangle. \\[.05in]
&= \frac{\eta}{16}\int\!dt\int\!du\int\!\frac{d\omega_S}{2\pi}\int\!\frac{d\omega_I}{2\pi}\int\!\frac{d\omega_S'}{2\pi}\int\!\frac{d\omega_I'}{2\pi}\,e^{i[(\omega_S-\omega_S')t - (\omega_I-\omega_I')u]}\left[1 + e^{-i[(\phi_S+\phi_I)-(\omega_S'-\omega_I')\Delta T]}\right. \nonumber \\[.05in]
&+ \left. e^{i[(\phi_S + \phi_I) - (\omega_S-\omega_I)\Delta T]} + e^{-i(\omega_S-\omega_I+\omega_I'-\omega_S')\Delta T} \right]\left\langle |\omega_S\rangle_S|\omega_I\rangle_I{}_I\langle \omega_I'|{}_S\langle \omega_S'|\right\rangle \\[.05in]
&= \frac{\eta}{8}\int\!\frac{d\omega_S}{2\pi}\int\!\frac{d\omega_I}{2\pi}\, \left[1+{\rm Re}\!\left(e^{i[(\phi_S+\phi_I)-(\omega_S-\omega_I)\Delta T]}\right)\right]\left\langle|\omega_S\rangle_S|\omega_I\rangle_I{}_I\langle \omega_I|{}_S\langle \omega_S|\right\rangle \\[.05in]
&= \frac{\eta}{8}\left[1 + {\rm Re}\!\left(e^{i(\phi_S+\phi_I)}\left\langle e^{-i(\hat{\omega}_S-\hat{\omega}_I)\Delta T}\right\rangle\right)\right],
\end{align}
where the second equality follows from Eqs.~(\ref{freqstateS}) and (\ref{freqstateI}), and the last equality follows from the source's emitting a single photon-pair.  Because the coincidence probability only depends on $\phi\equiv \phi_S+\phi_I$, we shall use the notation $P_{C_{\rm FI}}(\phi)$ in what follows.

The $0$-$\pi$ fringe visibility of a Franson interferometer with delay $\Delta T$ is defined as 
\begin{equation}
\label{eqFransonVisibility}
V_{\rm FI}(\Delta T) = \frac{P_{C_{\rm FI}}(0)-P_{C_{\rm FI}}(\pi)}{P_{C_{\rm FI}}(0)+P_{C_{\rm FI}}(\pi)} = {\rm Re}\!\left(\left\langle e^{-i(\hat{\omega}_S-\hat{\omega}_I)\Delta T}\right\rangle\right) 
= \langle \cos[(\hat{\omega}_S-\hat{\omega}_I)\Delta T]\rangle.
\end{equation}
The frequency-domain biphoton wave function associated with Eq.~(\ref{biphotonTime}),\begin{equation}
\Psi(\omega_S,\omega_I) = \frac{e^{-\omega_+^2\sigma^2_{\rm cor} - \omega_-^2\sigma^2_{\rm coh}}}{\sqrt{\pi/2\sigma_{\rm cor}\sigma_{\rm coh}}}
\label{biphotonFreq}
\end{equation}
where $\omega_+ \equiv (\omega_S+\omega_I)/2$ and $\omega_- = \omega_S -\omega_I$, makes 
\begin{align}
\langle\cos[(\hat{\omega}_S-\hat{\omega}_I)\Delta T]\rangle &= 
e^{-\langle(\hat{\omega}_S-\hat{\omega}_I)^2\rangle\Delta T^2/2} \\
& \le
1- \langle(\hat{\omega}_S-\hat{\omega}_I)^2\rangle\Delta T^2/2 + 
\langle(\hat{\omega}_S-\hat{\omega}_I)^2\rangle^2\Delta T^4/8\\
&= 1- \langle(\hat{\omega}_S-\hat{\omega}_I)^2\rangle\Delta T^2/2 + 
\langle(\hat{\omega}_S-\hat{\omega}_I)^4\rangle\Delta T^4/24,
\end{align}
with $\langle(\hat{\omega}_S-\hat{\omega}_I)^2\rangle = 1/4\sigma_{\rm coh}^2$ and the last equality following from Gaussian moment factoring \cite{Haykin}.  
In the presence of eavesdropping we use a three-term Taylor-series expansion to show that
\begin{equation}
\langle\cos[(\hat{\omega}_S-\hat{\omega}_I)\Delta T]\rangle \le 1 - \frac{\langle(\hat{\omega}_S-\hat{\omega}_I)^2\rangle\Delta T^2}{2}  
+ \frac{\langle(\hat{\omega}_S-\hat{\omega}_I)^4\rangle\Delta T^4}{24}.
\label{FIexpansion}
\end{equation}
Note that (\ref{FIexpansion}) does \em not\/\rm\ assume the biphoton has a Gaussian wave function, so it applies for any biphoton state---pure or mixed---of the signal and idler detected by Alice and Bob.
The third term in this inequality satisfies
\begin{equation}
\langle(\hat{\omega}_S-\hat{\omega}_I)^4\rangle\Delta T^4/24 \le 
\langle(\widetilde{\omega}_S-\widetilde{\omega}_I)^4\rangle\Delta T^4/24,
\label{omega4bound}
\end{equation}
where $\widetilde{\omega}_S$ and $\widetilde{\omega}_I$ are classical random variables obtained from frequency measurements at Alice and Bob's terminals respectively, i.e., by disabling the frequency-shifted arms in their conjugate Franson interferometers and relying on those interferometers' dispersive elements to make frequency information manifest in the observed arrival times \cite{footnote5}.  Using (\ref{omega4bound}) in (\ref{FIexpansion}) we get
\begin{equation}
\left<(\hat{\omega}_S-\hat{\omega}_I)^2\right> \leq \frac{2[1-V_\text{FI}(\Delta T)]}{\Delta T^2}+\frac{\langle (\widetilde{\omega}_S-\widetilde{\omega}_I)^4 \rangle\Delta T^2}{12}.
\label{Lemma1culm}
\end{equation}
Together with $\langle(\Delta\hat{\omega}_S-\Delta\hat{\omega}_I)^2\rangle \le \langle(\hat{\omega}_S-\hat{\omega}_I)^2\rangle$, (\ref{Lemma1culm}) completes the proof of Lemma~1.  

\section{Proof of Lemma 2}
The coincidence probability for our conjugate Franson interferometer when only a single photon-pair has been emitted by Alice's source is \cite{footnote6}
\begin{align}
P_{C_{\rm CFI}}(\phi_S,\phi_I) &= \frac{\eta}{16}\int\!dt\int_{t-T_g/2}^{t+T_g/2}\!du\int\!\frac{d\omega_S}{2\pi}\int\!\frac{d\omega_I}{2\pi}\int\!\frac{d\omega_S'}{2\pi}\int\!\frac{d\omega_I'}{2\pi}\, \left\langle\left[\hat{A}_S^\dagger(\omega_S) + e^{i\phi_S}\hat{A}_S^\dagger(\omega_S-\Delta\Omega) \right]\right.\nonumber\\ 
&\times \left[\hat{A}_I^\dagger(\omega_I) + e^{i\phi_I}\hat{A}_I^\dagger(\omega_I-\Delta\Omega) \right]\left[\hat{A}_S(\omega_S') + e^{-i\phi_S}\hat{A}_S(\omega_S'-\Delta\Omega) \right] \nonumber\\ 
&\times \left.\left[\hat{A}_I(\omega_I') + e^{-i\phi_I}\hat{A}_I(\omega_I'-\Delta\Omega) \right]\right\rangle e^{-i[\beta_2(\omega_S^2-\omega_I^2-\omega_S'^2+\omega_I'^2)/2 - (\omega_S-\omega_S')t +(\omega_I-\omega_I')u]}.
\label{CFI1}
\end{align}
Using Eq.~(\ref{Asingle}), this expression becomes
\begin{align}
P_{C_{\rm CFI}}(\phi_S,\phi_I) &= \frac{\eta}{16}\int\!dt\int_{t-T_g/2}^{t+T_g/2}\!du\int\!\frac{d\omega_S}{2\pi}\int\!\frac{d\omega_I}{2\pi}\int\!\frac{d\omega_S'}{2\pi}\int\!\frac{d\omega_I'}{2\pi}\, \left\langle\left[|\omega_S\rangle_S{}_S\langle 0| + e^{i\phi_S}|\omega_S-\Delta\Omega\rangle_S{}_S\langle 0| \right]\right.\nonumber\\ 
&\times \left[|\omega_I\rangle_I{}_I\langle 0| + e^{i\phi_I}|\omega_I-\Delta\Omega\rangle_I{}_I\langle 0| \right]\left[|0\rangle_S{}_S\langle\omega_S'| + e^{-i\phi_S}|0\rangle_S{}_S\langle\omega_S'-\Delta\Omega| \right] \nonumber\\ 
&\times \left.\left[|0\rangle_I{}_I\langle \omega_I'| + e^{-i\phi_I}|0\rangle_I{}_I\langle\omega_I'-\Delta\Omega| \right]\right\rangle e^{-i[\beta_2(\omega_S^2-\omega_I^2-\omega_S'^2+\omega_I'^2)/2 -(\omega_S-\omega_S')t + (\omega_I-\omega_I')u]}.
\label{CFI2}
\end{align}

Multiplying out the bracketed expressions inside the averaging leads to a sum of sixteen terms, but only four of them make nonzero contributions when $\Delta\Omega > T_g/2\beta_2 \gg 3/\sigma_{\rm coh}$.  To show that this is so, we begin by rewriting Eq.~(\ref{CFI2}) in terms of the sum and difference variables $t_+ \equiv (t+u)/2$, $t_-\equiv t-u$, $\omega_+\equiv (\omega_S+\omega_I)/2$, $\omega_-\equiv \omega_S-\omega_I$, $\omega'_+\equiv (\omega'_S+\omega'_I)/2$, and $\omega'_-\equiv \omega'_S-\omega'_I$, obtaining
\begin{align}
P_{C_{\rm CFI}}(\phi_S,\phi_I) &= \frac{\eta}{16}\int\!dt_+\int_{-T_g/2}^{T_g/2}\!dt_-\int\!\frac{d\omega_+}{2\pi}\int\!\frac{d\omega_-}{2\pi}\int\!\frac{d\omega'_+}{2\pi}\int\!\frac{d\omega'_-}{2\pi}\, \left\langle\left[|\omega_++\omega_-/2\rangle_S{}_S\langle 0| \right.\right.\nonumber \\
&\left.+\,\, e^{i\phi_S}|\omega_++\omega_-/2-\Delta\Omega\rangle_S{}_S\langle 0| \right]\ \left[|\omega_+-\omega_-/2\rangle_I{}_I\langle 0| + e^{i\phi_I}|\omega_+-\omega_-/2-\Delta\Omega\rangle_I{}_I\langle 0| \right] \nonumber \\
&\times \left[|0\rangle_S{}_S\langle\omega'_++\omega'_-/2| + e^{-i\phi_S}|0\rangle_S{}_S\langle\omega'_++\omega'_-/2-\Delta\Omega| \right] \nonumber\\ 
&\times \left.\left[|0\rangle_I{}_I\langle \omega'_+-\omega'_-/2| + e^{-i\phi_I}|0\rangle_I{}_I\langle\omega'_+-\omega'_-/2-\Delta\Omega| \right]\right\rangle \nonumber \\
&\times e^{-i[\beta_2(\omega_+\omega_--\omega'_+\omega'_-) -(\omega_+-\omega'_+)t_- - (\omega_--\omega'_-)t_+]}.
\label{CFI3}
\end{align}
Next, we perform the $t_-$, $t_+$, and $\omega'_-$ integrals and get
\begin{align}
P_{C_{\rm CFI}}(\phi_S,\phi_I) &= \frac{\eta}{16}\int\!\frac{d\omega_+}{2\pi}\int\!\frac{d\omega_-}{2\pi}\int\!\frac{d\omega'_+}{2\pi}\, \left\langle\left[|\omega_++\omega_-/2\rangle_S{}_S\langle 0| + e^{i\phi_S}|\omega_++\omega_-/2-\Delta\Omega\rangle_S{}_S\langle 0| \right]\right.\nonumber \\
& \times \left[|\omega_+-\omega_-/2\rangle_I{}_I\langle 0| + e^{i\phi_I}|\omega_+-\omega_-/2-\Delta\Omega\rangle_I{}_I\langle 0| \right] \nonumber \\
&\times \left[|0\rangle_S{}_S\langle\omega'_++\omega_-/2| + e^{-i\phi_S}|0\rangle_S{}_S\langle\omega'_++\omega_-/2-\Delta\Omega| \right] \nonumber\\ 
&\times \left.\left[|0\rangle_I{}_I\langle \omega'_+-\omega_-/2| + e^{-i\phi_I}|0\rangle_I{}_I\langle\omega'_+-\omega_-/2-\Delta\Omega| \right]\right\rangle \nonumber \\
&\times T_g\frac{\sin[(\omega_+-\omega'_+)T_g/2]}{(\omega_+-\omega'_+)T_g/2}
e^{-i\beta_2(\omega_+-\omega'_+)\omega_-}.
\label{CFI4}
\end{align}

Because the phase-matching bandwidth $B_{\rm PM}$ greatly exceeds $1/T_g$, the state-average term in 
Eq.~(\ref{CFI4}) is essentially unchanged for $\omega\equiv (\omega_++\omega'_+)/2$ excursions on the order of $10/T_g$.  Thus Eq.~(\ref{CFI4}) simplifies to
\begin{align}
P_{C_{\rm CFI}}(\phi_S,\phi_I) &= \frac{\eta}{16}\int\!\frac{d\omega}{2\pi}\int\!\frac{d\omega_-}{2\pi}\int\!\frac{d\omega'}{2\pi}\, \left\langle\left[|\omega+\omega_-/2\rangle_S{}_S\langle 0| + e^{i\phi_S}|\omega+\omega_-/2-\Delta\Omega\rangle_S{}_S\langle 0| \right]\right.\nonumber \\
& \times \left[|\omega-\omega_-/2\rangle_I{}_I\langle 0| + e^{i\phi_I}|\omega-\omega_-/2-\Delta\Omega\rangle_I{}_I\langle 0| \right] \nonumber \\
&\times \left[|0\rangle_S{}_S\langle\omega+\omega_-/2| + e^{-i\phi_S}|0\rangle_S{}_S\langle\omega+\omega_-/2-\Delta\Omega| \right] \nonumber\\ 
&\times \left.\left[|0\rangle_I{}_I\langle \omega-\omega_-/2| + e^{-i\phi_I}|0\rangle_I{}_I\langle\omega-\omega_-/2-\Delta\Omega| \right]\right\rangle \nonumber \\
&\times T_g\frac{\sin(\omega'T_g/2)}{\omega'T_g/2}
e^{-i\beta_2\omega'\omega_-}, 
\label{CFI5}
\end{align}
where $\omega'\equiv \omega_+-\omega'_+$.
Performing the $\omega'$ integral then gives us
\begin{align}
P_{C_{\rm CFI}}(\phi_S,\phi_I)&= \frac{\eta}{16}\int\!\frac{d\omega}{2\pi}\int_{-T_g/2\beta_2}^{T_g/2\beta_2}\!\frac{d\omega_-}{2\pi}\, \left\langle\left[|\omega+\omega_-/2\rangle_S{}_S\langle 0| + e^{i\phi_S}|\omega+\omega_-/2-\Delta\Omega\rangle_S{}_S\langle 0| \right]\right.\nonumber \\
& \times \left[|\omega-\omega_-/2\rangle_I{}_I\langle 0| + e^{i\phi_I}|\omega-\omega_-/2-\Delta\Omega\rangle_I{}_I\langle 0| \right] \nonumber \\
&\times \left[|0\rangle_S{}_S\langle\omega+\omega_-/2| + e^{-i\phi_S}|0\rangle_S{}_S\langle\omega+\omega_-/2-\Delta\Omega| \right] \nonumber\\ 
&\times \left.\left[|0\rangle_I{}_I\langle \omega-\omega_-/2| + e^{-i\phi_I}|0\rangle_I{}_I\langle\omega-\omega_-/2-\Delta\Omega| \right]\right\rangle.
\label{CFI6}
\end{align}

At this point we can quickly eliminate twelve terms from Eq.~(\ref{CFI6}).  In Eve's absence, any term containing $\langle |\omega_A\rangle_S{}_S\langle \omega_B|\otimes |\omega_A'\rangle_I{}_\langle \omega_B'|\rangle$ will only contribute if $|\omega_A-\omega_A'|\le 3/\sigma_{\rm coh}$ \em and\/\rm\ $|\omega_B-\omega_B'| \le 3/\sigma_{\rm coh}$.  Indeed, the magnitude of each of these noncontributor terms is at least $\exp(-\Delta \Omega^2\sigma^2_{\rm coh}/2)$ times smaller than the terms we will retain.  For our paper's system example, which assumes $\Delta \Omega/2\pi = 5$\,GHz, this attenuation factor is $\exp(-20)$ for $\sigma_{\rm coh} = 0.20\,$ns (corresponding to $T_f = 16\delta T$) and $\exp(-82)$ for $\sigma_{\rm coh} = 0.41\,$ns (corresponding to $T_f = 32\delta T$).   These numbers apply to detectors without timing jitter.  With timing jitter, the preceding attenuation factor becomes $\exp(-\Delta \Omega^2\beta_2^2\ln(2)/2\delta T^2)$, which equals $\exp(-8.98)$ for our $\Delta\Omega\beta_2 =  \sqrt{2}\,T_g$, $T_g = 108\,$ps , $\delta T = 30\,$ps system example.  It follows that the only terms which survive in $P_{C_{\rm CFI}}$ are as given below, where we have reverted to $\omega_S$, $\omega_I$ notation,
\begin{eqnarray}
\lefteqn{P_{C_{\rm CFI}}(\phi_S,\phi_I) = }\nonumber \\
&&\frac{\eta}{16}\int\!\frac{d\omega_S}{2\pi}\int_{\omega_S-T_g/2\beta_2}^{\omega_S+T_g/2\beta_2}\!\frac{d\omega_I}{2\pi}\, \left\langle\left[|\omega_S\rangle_S|\omega_I\rangle_I{}_I\langle \omega_I|{}_S\langle\omega_S| 
+e^{i(\phi_S + \phi_I)}|\omega_S-\Delta \Omega\rangle_S|\omega_I-\Delta\Omega\rangle_I{}_I\langle\omega_I|{}_S\langle \omega_S| \right.\right. \nonumber \\
&& \hspace{-.15in}\, + \left.\left. e^{-i(\phi_S+\phi_I)}|\omega_S\rangle_S|\omega_I\rangle_I{}_I\langle \omega_I-\Delta\Omega|{}_S\langle \omega_S-\Delta \Omega| + |\omega_S-\Delta\Omega\rangle_S|\omega_I-\Delta\Omega\rangle_I{}_I\langle \omega_I-\Delta\Omega|{}_S\langle \omega_S-\Delta\Omega|\right]\right\rangle.
\end{eqnarray}
The limits in the $\omega_I$ integral can be extended to $-\infty < \omega_-<\infty$, if no eavesdropping has occurred, because $\Delta\Omega > T_g/2\beta_2 \gg 3/\sigma_{\rm coh}$, in which case
\begin{eqnarray}
\lefteqn{P_{C_{\rm CFI}} = \frac{\eta}{8}\int\!\frac{d\omega_S}{2\pi}\int\!\frac{d\omega_I}{2\pi}\,\left[1+{\rm Re}\!\left(e^{i(\phi_S+\phi_I)}\langle|\omega_S-\Delta \Omega\rangle_S|\omega_I-\Delta\Omega\rangle_I{}_I\langle\omega_I|{}_S\langle \omega_S|\rangle\right)\right] }\\
&=& \frac{\eta}{8}\int\!\frac{d\omega_S}{2\pi}\int\!\frac{d\omega_I}{2\pi}\int\!dt_S\int\!dt_I\,\left[1+ {\rm Re}\!\left(e^{i(\phi_S+\phi_I)}\langle|\omega_S-\Delta \Omega\rangle_S|\omega_I-\Delta\Omega\rangle_I{}_I\langle t_I|{}_S\langle t_S|\rangle e^{i(\omega_St_S-\omega_It_I)} \right)\right] \hspace*{.2in}\\ 
&=& \frac{\eta}{8}\int\!dt_S\int\!dt_I\,\left[1+ {\rm Re}\!\left(e^{i(\phi_S+\phi_I)}e^{i(t_S-t_I)\Delta\Omega} \langle |t_S\rangle_S|t_I\rangle_I{}_I\langle t_I|{}_S\langle t_S|\rangle \right)\right] \\
&=& \frac{\eta}{8}\left[1+{\rm Re}\!\left(e^{i(\phi_S+\phi_I)}\left\langle e^{i(\hat{t}_S-\hat{t}_I)\Delta\Omega}\right\rangle\right)\right].\label{CFI7}
\end{eqnarray}
Now suppose that there is eavesdropping.  Any intrusion by Eve that degrades the frequency correlations to the point that the suppressed terms do contribute to $P_{C_{\rm CFI}}$ will be detected by the Franson interferometer and accounted for via Lemma~1, so we will complete our Lemma~2 proof using Eq.~(\ref{CFI7}).  Because the coincidence probability only depends on $\phi\equiv \phi_S + \phi_I$, we shall use the notation $P_{C_{\rm CFI}}(\phi)$ in what follows.

The conjugate-Franson interferometer's $0$-$\pi$ fringe visibility is defined to be
\begin{equation}
V_{\rm CFI}(\Delta \Omega)  = \frac{P_{C_{\rm CFI}}(0) - P_{C_{\rm CFI}}(\pi)}{P_{C_{\rm CFI}}(0) + P_{C_{\rm CFI}}(\pi)} = {\rm Re}\!\left(\left\langle e^{i(\hat{t}_S-\hat{t}_I)\Delta \Omega}\right\rangle\right) = 
\langle \cos[(\hat{t}_S-\hat{t}_I)\Delta \Omega]\rangle.
\end{equation}  
The time-domain biphoton wave function, Eq.~(\ref{biphotonTime}), makes
\begin{align}
\langle \cos[(\hat{t}_S-\hat{t}_I)\Delta \Omega]\rangle &= e^{-\langle (\hat{t}_S-\hat{t}_I)^2\rangle\Delta\Omega^2/2} \\
& \le 1- \langle (\hat{t}_S-\hat{t}_I)^2\rangle\Delta\Omega^2/2 + \langle(\hat{t}_S-\hat{t}_I)^2\rangle^2\Delta \Omega^4/8 \\ 
& = 1- \langle(\hat{t}_S-\hat{t}_I)^2\rangle\Delta\Omega^2/2 + 
\langle (\hat{t}_S-\hat{t}_I)^4\rangle\Delta \Omega^4/24,
\end{align}
with $\langle(\hat{t}_S-\hat{t}_I)^2\rangle = \sigma^2_{\rm cor}$ and the last equality again following from Gaussian moment factoring.  In the presence of eavesdropping we use a three-term Taylor-series expansion to show that 
\begin{equation}
\langle \cos[(\hat{t}_S-\hat{t}_I)\Delta \Omega]\rangle
\le 1 - \langle(\hat{t}_S-\hat{t}_I)^2\rangle\Delta\Omega^2/2 + 
\langle (\hat{t}_S-\hat{t}_I)^4\rangle\Delta \Omega^4/24.
\label{CFIexpansion}
\end{equation}
As noted in Lemma~1, (\ref{CFIexpansion}) does \em not\/\rm\ assume the biphoton has a Gaussian wave function, so it applies for any biphoton state---pure or mixed---of the signal and idler detected by Alice and Bob.
The third term in this inequality satisfies
\begin{equation}
\langle(\hat{t}_S-\hat{t}_I)^4\rangle\Delta \Omega^4/24 \le 
\langle(\widetilde{t}_S-\widetilde{t}_I)^4\rangle\Delta \Omega^4/24,
\label{t4bound}
\end{equation}
where $\widetilde{t}_S$ and $\widetilde{t}_I$ are classical random variables obtained from arrival-time measurements at Alice and Bob's terminals respectively, i.e., by disabling the long arms in their Franson interferometer's \cite{footnote6a}.  Using (\ref{t4bound}) in (\ref{CFIexpansion}) we get
\begin{equation}
\left<(\hat{t}_S-\hat{t}_I)^2\right> \leq \frac{2[1-V_\text{CFI}(\Delta \Omega)]}{\Delta \Omega^2}+\frac{\langle (\widetilde{t}_S-\widetilde{t}_I)^4 \rangle\Delta \Omega^2}{12}.
\label{Lemma2culm}
\end{equation}
Together with $\langle(\Delta\hat{t}_S-\Delta\hat{t}_I)^2\rangle \le \langle(\hat{t}_S-\hat{t}_I)^2\rangle$, (\ref{Lemma2culm}) completes the proof of Lemma~2.  

\section{Security calculations}
\subsection{The time-frequency covariance matrix}
The TFCM for the biphoton state from Eq.~(\ref{biphotonTime}) is 
\begin{equation}
\Gamma_0 = \left[ \begin{array}{ccc}
\gamma^0_{SS} & &\gamma^0_{SI}  \\[.1in]
\gamma^0_{IS} && \gamma^0_{II} 
\end{array} \right],
\end{equation}
where 
\begin{subequations}
\label{EqTFCM0}
\begin{align}
& \gamma^0_{SS} = \gamma^0_{II} = \left[\begin{array}{ccc}
\sigma^2_\text{cor}/4+\sigma^2_\text{coh} & &  0\\[.1in]
0 & &  1/4\sigma^2_\text{cor}+ 1/16\sigma^2_\text{coh}
\end{array}\right]\\[.1in]
& \gamma^0_{SI} = \gamma^0_{IS} = \left[\begin{array}{ccc}
-\sigma^2_\text{cor}/4+\sigma^2_\text{coh} & & 0\\[.1in]
0 & & 1/4\sigma^2_\text{cor} - 1/16\sigma^2_\text{coh}
\end{array}\right].
\end{align}
\end{subequations}
The root-mean-square two-photon correlation time is given by $\sigma_\text{cor} = \sqrt{2\ln(2)}/(2\pi B_{\rm PM})$ in terms of the FWHM phase-matching bandwidth (in Hz).  The root-mean-square two-photon pulse duration is given by $\sigma_\text{coh} = T_f/\sqrt{8\ln(2)}$ in terms of our protocol's frame duration---which is taken to be the FWHM coherence time---and is set by choice of the pump pulse's duration.  To achieve multiple secure bits per coincidence, we require $ \sigma_\text{coh} \gg \delta T/\sqrt{8\ln(2)} \gg \sigma_\text{cor} $. Under this condition, we have $ \langle \Delta \hat{t}_{S_0}^2\rangle = \langle \Delta \hat{t}_{I_0}^2 \rangle = \sigma^2_\text{coh} $, $ \langle ( \Delta \hat{t}_{S_0}-\Delta \hat{t}_{I_0})^2\rangle = \sigma^2_\text{cor} $, $ \langle \Delta\hat{\omega}_{S_0}^2\rangle = \langle \Delta\hat{\omega}_{I_0}^2\rangle = 1/4\sigma_\text{cor}^2  $, and $ \langle (\Delta \hat{\omega}_{S_0}-\Delta\hat{\omega}_{I_0})^2\rangle = 1/4\sigma_\text{coh}^2$, where the subscript 0 denotes the initial state produced by the source.  The Franson and conjugate-Franson interferometer's allow us to upper bound the variances of Alice and Bob's arrival-time and frequency differences, which we denote $ \langle (\Delta\hat{t}_S - \Delta\hat{t}_I)^2\rangle = (1+\xi_t) \langle (\Delta\hat{t}_{S_0}-\Delta\hat{t}_{I_0})^2\rangle$ and $ \langle (\Delta\hat{\omega}_S-\Delta\hat{\omega}_I)^2 \rangle = (1+\xi_\omega) \langle (\Delta\hat{\omega}_{S_0}-\Delta\hat{\omega}_{I_0})^2\rangle $, where $ \xi_t $ and $ \xi_\omega $ quantify the amount of excess noise.

In an operational system, $\xi_\omega $ and $ \xi_t $ will be bounded from the measured Franson and conjugate-Franson's $0$-$\pi$ visibilities, plus the fourth moments of the arrival-time and frequency differences obtained, respectively, from Franson and conjugate-Franson count records with one arm disabled at Alice and Bob's terminals.  For the theoretical assessment of the secure-key rate presented in the paper, we used assumed values for the $0$-$\pi$ visibilities and the following jitter-limited values for the fourth moments appearing in Lemmas~1 and 2:
\begin{equation}
\langle (\widetilde{\omega}_S-\widetilde{\omega}_I)^4\rangle = 3(\delta T/2\sqrt{\ln(2)}\beta_2)^4,
\end{equation} 
and
\begin{equation}
\langle (\widetilde{t}_S-\widetilde{t}_I)^4\rangle = 3(\delta T/2\sqrt{\ln(2)})^4.
\end{equation}
Here, we have assumed that our detectors have statistically independent, identically distributed Gaussian timing jitters whose FWHM $\delta T$ satisfies $ \delta T^2 \gg \beta_2^2\langle (\hat{\omega}_S-\hat{\omega}_I)^2\rangle$ and $ \delta T^2 \gg \langle (\hat{t}_S-\hat{t}_I)^2\rangle$.

\subsection{Off-diagonal elements in covariance sub-matrices}
The off-diagonal elements in  Eq.~(\ref{EqTFCM0}) are all zero, because Alice's source does not produce any time-frequency cross correlations. However, Eve's intrusion could result in a TFCM
\begin{equation}
\Gamma = \left[ \begin{array}{ccc}
\gamma_{SS} & &\gamma_{SI}  \\[.1in]
\gamma_{IS} && \gamma_{II} 
\end{array} \right],
\label{postpropTFCM}
\end{equation}
whose sub-matrices have nonzero off-diagonal elements.
Inasmuch as the Franson and conjugate-Franson interferometers do \em not\/\rm\ probe those off-diagonal elements, we now prove that the information they do provide---through Lemmas~1 and 2---suffices to upper bound Eve's Holevo information. 

We begin by introducing dimensionless time and frequency operators defined as follows
\begin{subequations}
\begin{align}
\hat{\tilde{t}}_J &= \frac{\hat{t}_J}{T}\\
\hat{\tilde{\omega}}_J &= \hat{\omega}_J T,
\end{align}
\end{subequations}
for $J = S,I$, where $ T \equiv \sqrt{2\sigma_\mathrm{coh}\sigma_\mathrm{cor}}$ is a normalization time that symmetrizes these conjugate observables, i.e., for Alice's SPDC source we have that $\langle \Delta\hat{\tilde{t}}_{J_0}^2\rangle = \langle \Delta\hat{\tilde{\omega}}_{J_0}^2\rangle$.   The biphoton wave function from Eq.~(\ref{biphotonTime}) now becomes a two-mode squeezed-vacuum state of the modes associated with the effective annihilation and creation operators
\begin{subequations}
\begin{align}
\hat{\tilde{a}}_J &= (\hat{\tilde{t}}_J - i\epsilon_J\hat{\tilde{\omega}}_J)/\sqrt{2} \\
\hat{\tilde{a}}^\dag_J &= (\hat{\tilde{t}}_J + i\epsilon_J\hat{\tilde{\omega}}_J)/\sqrt{2},
\end{align}
\end{subequations}
because $[\hat{\tilde{\omega}}_J,\hat{\tilde{t}}_K] = i\epsilon_J\delta_{JK}$ implies that 
$[\hat{\tilde{a}}_J,\hat{\tilde{a}}^\dagger_K] = \delta_{JK}$.
As a result, we can now apply security results of entanglement-based CVQKD to our TEE-based HDQKD protocol. 

Eve's arbitrary Gaussian attack can be decomposed into the following two steps:  Step~1, she interacts her ancillary state with the idler beam---while it is en route from Alice to Bob---in a manner that does not introduce any time-frequency off-diagonal elements in Alice and Bob's TFCM.  Step~2, she applies individual symplectic transformations to the idler state and her ancillary state, a process which creates such time-frequency off-diagonal elements \cite{serafini04}. Step~2 consists of a rotation followed by quadrature squeezing. Because it only involves local unitary operations, it does not affect Eve's Holevo information. Eve's goal here is to minimize $\delta\tilde{\omega}^2 \equiv \langle (\hat{\tilde{\omega}}_S-\hat{\tilde{\omega}}_I)^2 \rangle$,  and thus maximize Alice and Bob's 0-$\pi$ Franson-interferometer visibility so that they will underestimate the information her interaction has yielded \cite{footnote8}.  More specifically, Eve's goal in Step~2 is to make $\delta\tilde{\omega}^2$ smaller than $\delta\tilde{\omega}_{\rm diag}^2$, the 
$\langle (\hat{\tilde{\omega}}_S-\hat{\tilde{\omega}}_I)^2 \rangle$ value when she \em only\/\rm\ employs Step~1, i.e., when Alice and Bob's TFCM has no off-diagonal terms representing time-frequency cross correlations.

To see that Eve's effort in this regard is futile we proceed as follows, where, for notational simplicity, we have assumed that mean values have been subtracted out.  Let the effective idler annihilation operator after Step~1 be $\hat{\tilde{b}}$. The effective idler annihilation operator after Step~2 is then
\begin{equation}
\hat{\tilde{a}}_I = e^{i\varphi}\cosh(r) \hat{\tilde{b}} -e^{i(\theta-\varphi)}\sinh(r) \hat{\tilde{b}}^\dag,
\end{equation}
where $\varphi$ is the phase used in the rotation, $ r $ is the squeezing parameter, and $ \theta $ is the phase used in the squeezing operation. We now find that
\begin{equation}
\hat{\tilde{\omega}}_I = \left[\cos(\varphi) \cosh(r) + \cos(\theta-\varphi)\sinh(r)\right]\hat{\tilde{\omega}}_b + \left[\sin(\varphi) \cosh(r) - \sin(\theta-\varphi)\sinh(r)\right]\hat{\tilde{t}}_b,
\label{step2}
\end{equation}
from which the mean-squared frequency error is seen to obey
\begin{eqnarray}
\label{eqdeltaomega}
\delta\tilde{\omega}^2 &=& \langle \hat{\tilde{\omega}}_S^2\rangle - 2\left[\cos(\varphi) \cosh(r)+\cos(\theta-\varphi)\sinh(r)\right]\langle \hat{\tilde{\omega}}_S\hat{\tilde{\omega}}_b\rangle \notag\\
&+& \left[\cos(\varphi) \cosh(r)+\cos(\theta-\varphi)\sinh(r)\right]^2\langle \hat{\tilde{\omega}}_b^2 \rangle+\left[\sin(\varphi)\cosh(r)-\sin(\theta-\varphi)\sinh(r)\right]^2\langle\hat{\tilde{t}}_b^2\rangle,
\end{eqnarray}
where we have used the fact that Eve's Step~1 does not create any time-frequency cross correlations.  
Setting $\varphi=0$, $r=0$, and $\theta=0$ in this expression eliminates Eve's Step~2 and leads us to the benchmark value
\begin{equation} 
\delta\tilde{\omega}^2_{\rm diag} = \langle (\hat{\tilde{\omega}}_S - \hat{\tilde{\omega}}_b)^2\rangle,
\end{equation}
that she is trying to beat with her symplectic transformation.  Without loss of generality, we can assume that $\delta\tilde{\omega}^2_{\rm diag} \ll \langle \hat{\tilde{\omega}}_S^2\rangle$, because violating this condition makes Eve's presence extremely obvious to Alice and Bob.

To minimize $\delta\tilde{\omega}^2$, we set its partial derivatives with respect to $\varphi, r,$ and $\theta$ to zero.  For the partial derivative with respect to $\varphi$ we have
\begin{eqnarray}
\label{eqpartialphi}
\frac{\partial (\delta\tilde{\omega}^2)}{\partial \varphi} &=& 2\left[\sin(\varphi)\cosh(r) -\sin(\theta-\varphi)\sinh(r)\right]\langle \hat{\tilde{\omega}}_S\hat{\tilde{\omega}}_b\rangle \notag \\
&-& 2\left[\sin(\varphi) \cosh(r) - \sin(\theta -\varphi)\sinh(r)\right]\left[\cos(\varphi)\cosh(r) + \cos(\theta-\varphi)\sinh(r)\right](\langle \hat{\tilde{\omega}}_b^2 \rangle
-\langle \hat{\tilde{t}}_b^2\rangle),
\end{eqnarray}
which vanishes if 
\begin{equation}
\label{eqCondphi1}
\left[\sin(\varphi)\cosh(r) -\sin(\theta-\varphi)\sinh(r)\right] = 0 \tag{C.1}
\end{equation}
or
\begin{equation}
\label{eqCondphi2}
\langle \hat{\tilde{\omega}}_S\hat{\tilde{\omega}}_b\rangle = \left[\cos(\varphi)\cosh(r) + \cos(\theta-\varphi)\sinh(r) \right](\langle\hat{\tilde{\omega}}^2_b\rangle-\langle \hat{\tilde{t}}^2_b\rangle) \tag{C.2}.
\end{equation}
Condition~(\ref{eqCondphi1}), combined with the fact that Eve's Step~1 does not create any time-frequency cross correlations, implies that the TFCM will not have any time-frequency off-diagonal terms.  Hence we will only carry forward Condition~(\ref{eqCondphi2}) in trying to see if Eve's Step~2 helps her.  

For the partial derivative with respect to $\theta$ we have
\begin{eqnarray}
\label{eqpartialtheta}
\frac{\partial (\delta\tilde{\omega}^2)}{\partial \theta} &=& 2\sin(\theta-\varphi)\sinh(r)\langle \hat{\tilde{\omega}}_S\hat{\tilde{\omega}}_b\rangle-2\left[\cos(\varphi)\cosh(r) +\cos(\theta-\varphi)\sinh(r)\right]\left[\sin(\theta-\varphi)\sinh(r)\right]\langle\hat{\tilde{\omega}}^2_b\rangle\notag\\
&-& 2\left[\sin(\varphi) \cosh(r)-\sin(\theta-\varphi)\sinh(r)\right]\cos(\theta-\varphi)\sinh(r)\langle \hat{\tilde{t}}^2_b\rangle.
\end{eqnarray}
If the $\varphi$ partial derivative vanishes because (\ref{eqCondphi2}) is satisfied, we need
\begin{equation}
\label{eqCondphi2theta1}
\sin(\theta) = 0 \tag{C.2.1}
\end{equation}
or
\begin{equation}
\label{eqCondphi2theta2}
r = 0 \tag{C.2.2}
\end{equation}
to make the $\theta$ partial derivative vanish.
 
For the partial derivative with respect to $r$ we have
\begin{eqnarray}
\label{eqpartialr}
\frac{\partial (\delta\tilde{\omega}^2)}{\partial r} &=& -2\left[\cos(\varphi) \sinh(r) + \cos(\theta - \varphi)\cosh(r)\right]\langle\hat{\tilde{\omega}}_S\hat{\tilde{\omega}}_b \rangle\notag\\
&+&2\left[\cos(\varphi)\cosh(r) + \cos(\theta-\varphi)\sinh(r)\right]\left[\cos(\varphi)\sinh(r)+\cos(\theta-\varphi)\cosh(r)\right]\langle\hat{\tilde{\omega}}^2_b\rangle\notag\\
&+&2\left[\sin(\varphi)\cosh(r)-\sin(\theta-\varphi)\sinh(r)\right]\left[\sin(\varphi) \sinh(r) - \sin(\theta-\varphi)\cosh(r)\right]\langle\hat{\tilde{t}}^2_b \rangle.
\end{eqnarray}
If the $\varphi$ and $\theta$ partial derivatives vanish because Conditions~(\ref{eqCondphi2}) and (\ref{eqCondphi2theta1}) are satisfied, then the $r$ partial derivative equals zero when $\sinh(r) = \pm \cosh(r)$, where the plus sign applies for $\cos(\theta) = -1$ and the minus sign for $\cos(\theta) = 1$.  We then find that $\delta\tilde{\omega}^2 = \langle \hat{\tilde{\omega}}_S^2\rangle \gg \delta\tilde{\omega}^2_{\rm diag}$, making Eve's presence very detectable if she elects to perform Step~2 with these parameter values.  

At this point, the only remaining case that might provide Eve some utility from her Step~2 is when Conditions~(\ref{eqCondphi2}) and (\ref{eqCondphi2theta2}) are satisfied and the $r$ partial derivative vanishes.  Here, Step~2 corresponds to rotation without squeezing, in which case we need $\cos(\theta) = 0$ to make the $r$ partial derivative vanish, yielding
\begin{equation}
\delta\tilde{\omega}^2 = \langle \hat{\tilde{\omega}}_S^2\rangle - \cos^2(\varphi)(\langle \hat{\tilde{\omega}}_b^2\rangle - \langle \hat{\tilde{t}}_b^2\rangle) + \langle \hat{\tilde{t}}_b^2\rangle.
\end{equation}
If $\langle \hat{\tilde{\omega}}_b^2\rangle = \langle \hat{\tilde{t}}_b^2\rangle$, we get $\delta\tilde{\omega}^2 \gg \delta\tilde{\omega}^2_{\rm diag}$, so Eve shouldn't use a symplectic transformation after her Step~1.  If $\langle \hat{\tilde{\omega}}_b^2\rangle > \langle \hat{\tilde{t}}_b^2\rangle$, then her $\delta\tilde{\omega}^2$ is minimized by choosing $\cos^2(\varphi) = 1$, and Eq.~(\ref{step2}) shows that no off-diagonal TFCM elements arise from time-frequency cross correlations.  If $\langle \hat{\tilde{\omega}}_b^2\rangle < \langle \hat{\tilde{t}}_b^2\rangle$, then Eve's $\delta\tilde{\omega}^2$ is minimized by choosing $\cos(\varphi) = 0$, and we get $\delta\tilde{\omega}^2 = \langle \hat{\tilde{\omega}}_S^2\rangle + \langle \hat{\tilde{t}}_b^2\rangle \gg \delta\tilde{\omega}^2_{\rm diag}$, making her presence very detectable if she elects to perform Step~2 with these parameter values.  

To summarize, we have shown that for any value of Eve's Holevo information, the minimum value of $\delta\tilde{\omega}^2$ can be achieved \em without\/\rm\ introducing any off-diagonal elements into the TFCM sub-matrices \cite{footnote8}.  Hence Lemmas~1 and 2 suffice to bound the information that Eve obtains from an optimized collective attack.

\subsection{Alice and Bob's Shannon Information}
Alice and Bob postselect frames in which each of them had \textit{at least} one detection, which may be either a photon detection or a dark count. A fraction $q$ of these postselected frames are reconciled to generate key, while the rest are used for the Franson and conjugate-Franson measurements needed to estimate $\xi_t$ and $\xi_\omega$ and the decoy-state measurements needed for parameter estimation.  The probability for Alice and Bob to postselect a frame is therefore
\begin{equation}
p_r = \sum_{n = 0}^\infty p_s(n)\left[1-(1-\eta_A)^n(1-p_d)\right]\left[1-(1-\eta_B\eta_P)^n(1-p_d)\right],
\end{equation}
where: $p_s(n) = \mu^ne^{-\mu}/n!$ gives the probability that Alice's source emits an $n$-pair state in terms of the average number, $\mu$, of pairs emitted per frame \cite{footnote9}; $\eta_A$  and $\eta_B$ are Alice and Bob's detection efficiencies;  $\eta_P$ is the transmissivity of the fiber-propagation link from Alice's source to Bob' terminal; and $p_d$ is the probability of one dark-count occurring in a frame. Note that we are neglecting the possibility of multiple dark counts occurring in a frame, because the product of the frame duration and the dark-count rate (for typical SNSPDs) is much smaller than one.
 
When either Alice or Bob register more than one detection in a frame, they discard those data and randomly choose an arrival time from a Gaussian distribution whose variance equals their terminal's TFCM entry for arrival time plus the timing-jitter variance \cite{footnote10}. It follows that there are five possibilities for Alice and Bob's joint arrival-time distribution, from which their Shannon information can be found.
\begin{enumerate}
\item Their arrival times are jointly Gaussian random variables with a covariance matrix that is a submatrix of the TFCM augmented to account for their detectors' timing jitters. This case is a postselected frame in which Alice's source emitted one photon-pair and neither Alice nor Bob had a dark count.
\item Their arrival times are independent Gaussian random variables with variances equal to the values from the TFCM plus the variance of their detectors' timing jitters. This case is a postselected frame in which one of two situations occurred:  (1)  Alice's source emitted multiple photon-pairs and both Alice and Bob registered at least one photon detection; or (2)  Alice's source emitted one photon-pair and both Alice and Bob registered photon detections with at least one of them also having a dark count.  (Strictly speaking, there could be some correlation between Alice and Bob's arrival times in this case.  By neglecting such a possibility, we are underestimating its contribution to their Shannon information.)  
\item Alice's arrival time is a Gaussian random variable with variance equal to the value from the TFCM plus the variance of her detector's timing jitter, and Bob's is uniformly distributed over the frame. This case is a postselected frame in which Alice registered at least one photon detection and Bob had a dark count without a photon detection.
\item Alice's arrival time is uniformly distributed over the frame, and Bob's is a Gaussian random variable with variance equal to the value from the TFCM plus the variance of his detector's timing jitter.  This case is a postselected frame in which Alice had a dark count without a photon detection and Bob registered at least one photon detection.
\item Both Alice's and Bob's arrival times are uniformly distributed over the frame. This case is a postselected frame in which Alice and Bob both had dark counts and neither had a photon detection.
\end{enumerate}
Given that a particular frame has been postselected, the conditional occurrence probabilities for the preceding five events are
\begin{subequations}
\begin{align}
P_1 &=  p_s(1)\eta_A\eta_B\eta_P(1-p_d)^2/p_r\\
P_2 &= \sum_{n = 2}^\infty p_s(n)\left[1-(1-\eta_A)^n\right]\left[1-(1-\eta_B\eta_P)^n\right]/p_r + p_s(1)\eta_A\eta_B\eta_P(2p_d-p_d^2)/p_r\\
P_3 &= \sum_{n = 1}^\infty p_s(n)\left[1-(1-\eta_A)^n\right]\left[p_d(1-\eta_B\eta_P)^n\right]/p_r\\
P_4 &= \sum_{n = 1}^\infty p_s(n)\left[p_d(1-\eta_A)^n\right]\left[1-(1-\eta_B\eta_P)^n\right]/p_r\\
P_5 &= \sum_{n = 0}^\infty p_s(n)p_d^2(1-\eta_A)^n(1-\eta_B\eta_P)^n/p_r.
\end{align}
\end{subequations}

Alice and Bob's Shannon information is given by
\begin{equation}
I(A;B) = \int\!dt_Adt_B\,p_{T_A,T_B}(t_A,t_B)\log_2\left(\frac{p_{T_A,T_B}(t_A,t_B)}{p_{T_A}(t_A)p_{T_B}(t_B)}\right),
\end{equation}
for which we need the joint probability density function, $p_{T_A,T_B}(t_A,t_B)$, for their measured arrival times, $T_A$ and $T_B$, from which the marginal densities, $p_{T_A}(t_A)$ and $p_{T_B}(t_B)$, are easily obtained.  The mean values of $T_A$ and $T_B$ are invariant to which of the preceding five postselection events has occurred.  Thus they do not affect Alice and Bob's Shannon information, and so we will set them to zero.

The joint density function we need can be found from
\begin{equation}
p_{T_A,T_B}(t_A,t_B) = \sum_{i=1}^5P_i \, p_{T_A,T_B\mid i}(\,t_A,t_B\mid i\,),
\end{equation}
where $p_{T_A,T_B\mid i}(\,t_A,t_B\mid i\,)$ is the joint probability density for $T_A$ and $T_B$ given that event $i$ has occurred.  The conditional probability densities for events 2 through 5 are easily found, because $T_A$ and $T_B$ are statistically independent given that one of these events has occurred.  We have that
\begin{subequations}
\begin{align}
p_{T_A,T_B\mid 2}(\,t_A,t_B\mid 2\,) &=  p_G(t_A; \sigma_A^2)\,p_G(t_B;\sigma_B^2)\\ 
p_{T_A,T_B\mid 3}(\,t_A,t_B\mid 3\,) &=  p_G(t_A;\sigma_A^2)\,p_U(t_B; T_f)\\
p_{T_A,T_B\mid 4}(\,t_A,t_B\mid 4\,) &=  p_U(t_A;T_f)\,p_G(t_B;\sigma_B^2)\\ 
p_{T_A,T_B\mid 2}(\,t_A,t_B\mid 5\,) &=  p_U(t_A;T_f)\,p_U(t_B;T_f),
\end{align}
\end{subequations}
where $p_G(t;\sigma^2)$ is a Gaussian probability density with zero mean and variance $\sigma^2$, $p_U(t;T_f)$ is a uniform probability density over the interval of $[-T_f/2,T_f/2]$, and 
\begin{subequations}
\begin{align}
\sigma_A^2 &= \langle \Delta \hat{t}_S^2\rangle + (\delta T/2.35)^2\\ 
\sigma_B^2 &= \langle \Delta \hat{t}_I^2\rangle + (\delta T/2.35)^2.
\end{align}
\end{subequations}
When event~1 has occurred, $T_A$ and $T_B$ are jointly-Gaussian random variables with zero means and covariance matrix
\begin{equation}
\Lambda = \left[\begin{array}{ccc}
\sigma_A^2 & & \langle \Delta \hat{t}_S\Delta \hat{t}_I\rangle \\[.05in]
\langle \Delta \hat{t}_S\Delta \hat{t}_I\rangle && \sigma_B^2\end{array}
\right].
\end{equation}

\subsection{Eve's Holevo information}
Here we describe how Lemmas~1 and 2 permit us to place an upper bound on Eve's Holevo information.  Alice and Bob's TFCM has the form given in Eq.~(\ref{postpropTFCM}), with
\begin{subequations}
\label{submatTFCM}
\begin{align}
& \gamma_{SS} =\gamma_{SS}^0\\[.1in]
& \gamma_{SI} = \gamma_{IS} = \left[\begin{array}{ccc}
1-\eta_t & &  0\\[.1in]
0 & & 1-\eta_\omega\end{array}\right]\gamma^0_{SI}\\[.1in]
&\gamma_{II} = \left[\begin{array}{ccc}
1+\epsilon_t & &  0\\[.1in]
0 & & 1+\epsilon_\omega\end{array}\right]\gamma^0_{II},
\end{align}
\end{subequations}
where $\{\eta_t$, $\eta_\omega\}$ quantifies signal-idler correlation loss, and $\{\epsilon_t,\epsilon_\omega\}$ quantifies idler excess noise.  By means of Lemmas~1 and 2, the Franson and conjugate-Franson measurements provide upper bounds on the mean-squared arrival-time and frequency differences giving us values for the the excess-noise factors $ \xi_t $ and $ \xi_\omega $.  These values do not, however, determine $\{\eta_t,\eta_\omega\}$ and $\{\epsilon_t,\epsilon_\omega\}$.   Nevertheless, knowing $ \xi_t $ and $ \xi_\omega $, restricts the set, $\mathcal{M} $, of physically-allowed TFCMs.  Our upper bound on Eve's Holevo information is the maximum of her Holevo information over the TFCMs in $\mathcal{M}$.   

For a given a TFCM a Gaussian attack maximizes Eve's Holevo information. Thus, we can assume that the Alice, Bob, and Eve share a pure Gaussian state in evaluating that Holevo information.  We have that her Holevo information for covariance matrix $ \Gamma $ is
\begin{equation}
\label{eqChiAE}
\chi_\Gamma(A;E) = S(\hat{\rho}_E) - \int\!dt\, p_{T_A}(t_A) S(\hat{\rho}_{E|T_A = t_A}),
\end{equation}
where $S(\hat{\rho}) = - {\rm Tr}[\hat{\rho}\log_2(\hat{\rho})]$ is the von Neumann entropy of the state $\hat{\rho}$. Because Alice, Bob, and Eve's joint quantum state is pure, we have $S(\hat{\rho}_E) = S(\hat{\rho}_{AB})$. Conditioned on Alice's measurement, the quantum state shared by Bob and Eve is also pure, so that $S(\hat{\rho}_{E|T_A = t_A}) = S(\hat{\rho}_{B|T_A = t_A})$. Furthermore, because all these states are Gaussian, the von Neumann entropy of Bob and Eve's conditional quantum state is independent of Alice's measurement result. Thus, we can drop the integral in Eq.~(\ref{eqChiAE}) and get
\begin{equation}
\label{eqChiAE_final}
\chi_\Gamma(A;E) = S(\hat{\rho}_{AB})-S(\hat{\rho}_{B|T_A}).
\end{equation}

To evaluate $\chi_\Gamma(A;E)$ from Eq.~(\ref{eqChiAE_final}), we define
\begin{subequations}
\begin{align}
I_1 &= \langle\Delta\hat{t}_S^2\rangle\langle \Delta\hat{\omega}_S^2\rangle\\
I_2 &= \langle\Delta\hat{t}_I^2\rangle\langle \Delta\hat{\omega}_I^2\rangle\\
I_3 &= \langle\Delta\hat{t}_S\Delta\hat{t}_I\rangle \langle\Delta\hat{\omega}_S\Delta\hat{\omega}_I\rangle \\
I_4 &= (\langle\Delta\hat{t}_S^2\rangle\langle \Delta\hat{t}_I^2\rangle - \langle\Delta\hat{t}_S\Delta\hat{t}_I\rangle^2)(\langle\Delta\hat{\omega}_S^2\rangle\langle \Delta\hat{\omega}_I^2\rangle - \langle\Delta\hat{\omega}_S\Delta\hat{\omega}_I\rangle^2)\\
d_{\pm} &= \frac{1}{\sqrt{2}}\sqrt{I_1+I_2+2I_3\pm\sqrt{(I_1+I_2+2I_3)^2-4I_4}}.
\end{align}
\end{subequations}
We then have that $S(\hat{\rho}_{AB}) = f(d_+)+f(d_-) $, where 
\begin{equation}
f(d) = (d+1/2)\log_2(d+1/2)-(d-1/2)\log_2(d-1/2).
\end{equation}
and
\begin{equation}
S(\rho_{B|t}) = f\!\left(\sqrt{\det[\gamma_{I|T_A}]}\right),
\end{equation}
where Bob's conditional covariance matrix is
\begin{equation}
 \gamma_{I|T_A} = \left[\begin{array}{ccc}
 \langle\Delta\hat{t}_I^2\rangle - \langle\Delta\hat{t}_S\Delta\hat{t}_I\rangle^2/\langle\Delta\hat{t}_S^2\rangle & & 0\\[.1in]
 0 & & \langle \Delta\hat{\omega}_I^2\rangle
 \end{array}
 \right].
 \end{equation}
Our upper bound on Eve's Holevo information is now found from 
\begin{equation}
\chi^\text{UB}_{\xi_t,\xi_\omega}(A;E) = \sup_{\Gamma\in \mathcal{M} }\{\chi_{\Gamma}(A;E)\}.
\end{equation}

\subsection{Secure-key rate}
For a postselected frame that originated from emission of one photon-pair, Eve's Holevo information about Alice's measurement result is at most $ \chi^\text{UB}_{\xi_t,\xi_\omega}(A;E) $. For a postselected frame that originated from emission of multiple photon-pairs, we grant Eve perfect information about Alice's measurement result. The fraction of the postselected frames that originated from emission of one photon-pair is
\begin{equation}
F = p_s(1)\left[1-(1-\eta_A)(1-p_d)\right]\left[1-(1-\eta_B\eta_P)(1-p_d)\right]/p_r.
\end{equation}
The error-correcting code used during reconciliation to establish $n_R$ shared random bits between Alice and Bob employs, on average, $ n_\text{ECC} $ syndrome bits, where 
\begin{equation}
n_R = \beta I(A;B) + n_\text{ECC},
\end{equation}
with $0 <  \beta \le 1 $ being the code's reconciliation efficiency \cite{footnote11}.  In total, therefore, Eve will have captured at most $ (1-F)n_R+n_\text{ECC}+F\chi^\text{UB}_{\xi_t,\xi_\omega}(A;E) $ bits of information per postselected frame. The secure-key rate, in bits per postselected frame, is the difference between $n_R$ and Eve's total information, hence it satisfies
\begin{equation}
\Delta I(A;B) \ge \beta I(A;B) - (1-F)n_R - F\chi^\text{UB}_{\xi_t,\xi_\omega}(A;E).
\end{equation}
When all postselected frames originated from emissions of one photon-pair, i.e., $F=1$, we recover CVQKD's secure-key rate bound, $ \Delta I(A;B) \ge \beta I(A;B) - \chi^{UB}_{\xi_t,\xi_\omega}(A;E)$, showing that, as expected, the emission of multiple photon-pairs reduces the secure-key rate of time-energy entanglement based HDQKD. The secure-key rate in bits per second for our time-energy entanglement based HDQKD protocol is the product of the frame postselection rate and the secure-key rate in bits per postselected frame, viz., 
\begin{equation}
\text{SKR} \geq \frac{q p_r}{3T_f}\left[\beta I(A;B)-(1-F)n_R- F \chi^\text{UB}_{\xi_t,\xi_\omega}(A;E)\right],
\end{equation}
where we have accounted for frames occurring once every $3T_f$ seconds.


\begin{thebibliography}{}

\bibitem{gisin02}
N. Gisin, G. Ribordy, W. Tittel, and H. Zbinden, \rmp \textbf{74,} 145--195 (2002).

\bibitem{grosshans03}
F. Grosshans, G. Van Assche, J. Wenger, R. Brouri, N. J. Cerf, and
P. Grangier,  \nat \textbf{421,} 238--241 (2003).

\bibitem{lodewyck07}
J. Lodewyck, \em et al\/\rm., \pra \textbf{76,} 042305
(2007).

\bibitem{bennett84}
C. H. Bennett and G. Brassard, in Proceedings of IEEE International Conference on Computers, Systems, and Signal Processing, Bangalore, India (IEEE, New York, 1984) p.~175--179.

\bibitem{ekert91}
A. K. Ekert, \prl \textbf{67,} 661--663 (1991).

\bibitem{xuan09}
Q. D. Xuan, Z. Zhang, and P. L. Voss, \opex \textbf{17,} 24244--24249 (2009).

\bibitem{jouguet13}
P. Jouguet, S. Kunz-Jacques, A. Leverrier,	P. Grangier, and E. Diamanti, Nature Photon. \textbf{7,} 378--381 (2013).

\bibitem{wang05}
X.-B. Wang, \prl \textbf{94,} 230503 (2005).

\bibitem{lo05}
H.-K Lo, X. F. Ma, and K. Chen, \prl \textbf{94,} 230504 (2005).

\bibitem{schmitt-manderbach07}
T. Schmitt-Manderbach, \em et al\/\rm., \prl \textbf{98,} 010504 (2007).

\bibitem{rosenberg07}
D. Rosenberg, \em et al\/\rm., \prl \textbf{98,} 010503 (2007).

\bibitem{zhang08}
L. Zhang, C. Silberhorn, I. A. Walmsley, \prl \textbf{100,} 110504 (2008).

\bibitem{tittel00}
W. Tittel, J. Brendel, H. Zbinden, and N. Gisin, \prl \textbf{84,} 4737--4740 (2000).

\bibitem{ali-khan07}
I. Ali-Khan, C. J. Broadbent, and J. C. Howell, \prl \textbf{98,} 060503 (2007).

\bibitem{nunn13}
J. Nunn, L. J. Wright, C. S\"{o}ller, L. Zhang, I. A. Walmsley and B. J. Smith, \opex {\bf 21,} 15959 (2013).

\bibitem{navascues06}
M. Navascu\'{e}s, F. Grosshans, and A. Ac\'{i}n, \prl \textbf{97,} 190502 (2006).

\bibitem{garcia06}
R. Garc\'{i}a-Patr\'{o}n and N. J. Cerf, \prl \textbf{97,} 190503 (2006).

\bibitem{mower12}
J. Mower, Z. Zhang, P. Desjardins, C. Lee, J. H. Shapiro, and D. Englund, \pra \textbf{87,} 062322 (2013).

\bibitem{mower11}J. Mower, F. N. C. Wong, J. H. Shapiro, and D. Englund, arXiv:1110.4867 [quant-ph].

\bibitem{brougham13}
T. Brougham, S. M. Barnett, K. T. McCusker, P. G. Kwiat, and D. J. Gauthier, J. Phys. B:  Atom. Mol. Opt. Phys. \textbf{46,} 104010 (2013).

\bibitem{grice10}A similar interferometer was suggested in W. Grice, \em et al\/\rm., \em Digest of Frontiers in Optics 2010\/\rm\ (Opt. Soc. of America, Washington, DC, 2010) paper~FTuG3.

\bibitem{footnote0}We take $T_f$ to be the full-width at half-maximum coherence time of the SPDC source and put $T_f$-sec-long buffer intervals on both sides of each frame to minimize the likelihood of more than one pair per frame being emitted.  There will be many empty frames because the average number of pairs per frame will be much smaller than one.

\bibitem{footnote1}The SPDC source's pair flux will be kept well below 1\,pair per $T_f$\,sec frame.  Nevertheless, security demands that multiple-pair frames be accounted for.  Later we will describe how that will be done via decoy states.

\bibitem{footnote2}Alice and Bob's clocks must be synchronized to better than their detectors' timing jitters, but this will not be problematic.  Synchronization of remote optical clocks is an active research area that focuses on femtosecond and sub-femtosecond precision, which is far better than the $\sim$10\,ps we require, see, e.g., J. Kim, J. A. Cox, J. Chen, and F. X. K\"{a}rtner, Nature Photon. {\bf 2,} 733 (2008), and K. Predehl \em et al\/\rm., Science {\bf 336,} 441 (2012).  Furthermore, although Eve's manipulation of Alice and Bob's timing channel could degrade their Shannon information, any information it affords Eve about their raw key is still bounded above by $\chi^{\rm UB}_{\xi_t,\xi_\omega}(A;E)$.

\bibitem{Dixon13}P. B. Dixon, J. H. Shapiro, and F. N. C. Wong, \opex {\bf 21,} 5879 (2013).

\bibitem{suppl}
See the Supplemental Material for the proof.

\bibitem{wolf06}
M. M. Wolf, G. Giedke, and J. I. Cirac, \prl \textbf{96,} 080502 (2006).

\bibitem{franson89}
J. D. Franson, \prl \textbf{62,} 2205--2208 (1989).

\bibitem{footnote3}The various time intervals associated with our protocol satisfy
$T_f = \sqrt{8\ln(2)}\sigma_{\rm coh} > \Delta T > T_g > \delta T \gg \sigma_{\rm cor}.$

\bibitem{footnote4}Alice and Bob use all frames in which they have coincidences.  The Supplemental Material explains how they deal with frames that contain multiple coincidences.

\bibitem{gottesman04}
D. Gottesman, H.-K. Lo, N. L\"{u}tkenhaus, and J. Preskill, Quantum Inf. Comput. \textbf{4,} 325--360 (2004).

\bibitem{zhong13}T. Zhong and F. N. C. Wong, Phys. Rev. A {\bf 88,} 020103(R) (2013).

\end{thebibliography}

\begin{thebibliography}{}

\bibitem{YuenShapiro}H. P. Yuen and J. H. Shapiro, IEEE Trans. Inform. Theory {\bf 24,} 657 (1978).

\bibitem{Blow}K. J. Blow, R. Loudon, and S. J. D. Phoenix, Phys. Rev A {\bf 42,} 4102 (1990).

\bibitem{Shapiro}J. H. Shapiro, IEEE J. Sel. Top. Quantum Electron. {\bf 15,} 1547 (2009). 


\bibitem{footnote1}
The state $|\omega\rangle_S$ ($|\omega\rangle_I$) represents a single photon of the signal (idler) at frequency $\omega_P/2 + \omega$ ($\omega_P/2-\omega$).

\bibitem{footnote2}
We are neglecting dark counts here.  A dark-count occurrence will reduce the interferometer's fringe visibility, leading to a higher inferred value, from Lemma~1, for $\langle (\Delta\hat{\omega}_S-\Delta\hat{\omega}_I)^2\rangle$.

\bibitem{footnote3}We are assuming perfectly balanced interferometers at Alice and Bob's terminals, which can always be accomplished by purposeful attenuation in the less lossy arms.  

\bibitem{footnote4}Here, without loss of generality, we have suppressed the propagation delays from the SPDC source to Alice and Bob's detectors.    

\bibitem{Haykin}S. Haykin, \em Adaptive Filter Theory, 2nd edition\/\rm\ (Prentice Hall, Englewood
Cliffs, 1991).

\bibitem{footnote5}It would be more efficient to eliminate those interferometers' beam splitters as well as their frequency-shifted arms.  

\bibitem{footnote6}
We are neglecting dark counts here.  A dark-count occurrence will reduce the interferometer's fringe visibility leading to a higher inferred value, from Lemma~2, for $\langle (\Delta\hat{t}_S-\Delta\hat{t}_I)^2\rangle$.

\bibitem{footnote6a}It would be more efficient to eliminate those interferometers' beam splitters as well as their long arms. 

\bibitem{serafini04}
A. Serafini, F. Illuminati and S. De Siena, J. Phys. B: At. Mol. Opt. Phys. \textbf{37,} L21-L28 (2004).

\bibitem{footnote8}Eve could also aim to use off-diagonal elements to minimize $\langle(\hat{\tilde{t}}_S-\hat{\tilde{t}}_I)^2\rangle$ to maximize Alice and Bob's 0-$\pi$ conjugate-Franson visibility.  We will prove Eve derives no benefit from minimizing $\delta\tilde{\omega}^2$ with her symplectic transformation.  A similar argument will prove that the same is true if she tries, instead, to minimize $\delta\tilde{t}^2\equiv \langle(\hat{\tilde{t}}_S-\hat{\tilde{t}}_I)^2\rangle$ with that transformation.  Because we will consider security when only a Franson interferometer is employed---as well as when both interferometers are used---we focus our attention on the  case of 
$\delta\tilde{\omega}^2$ minimization.

\bibitem{footnote9}We are using the Poisson distribution here because it is an excellent approximation to the exact statistics when the product of the pump pulse's $\sim$ns duration and the SPDC crystal's $\sim$THz phase-matching bandwidth is much greater than one.

\bibitem{footnote10}This procedure may not be optimal, but it enables us to more easily calculate Alice and Bob's Shannon information.

\bibitem{footnote11}Our reconciliation procedure allows Alice and Bob to share an arbitrarily chosen number of bits, $n_R$, as is the case in CVQKD.  However, all such bits in excess of $\beta I(A;B)$ are obtained via public communication through the error-correction code they employ.  Consequently, Eve is able to capture $n_{\rm ECC} = n_R - \beta I(A;B)$ bits from that public-channel reconciliation procedure.


\end{thebibliography}
\end{document}